\documentclass[sigconf]{acmart}
\AtBeginDocument{%
  }

\setcopyright{acmlicensed}
\copyrightyear{2018}
\acmYear{2018}
\acmDOI{XXXXXXX.XXXXXXX}
\acmConference[Conference acronym 'XX]{Make sure to enter the correct
  conference title from your rights confirmation email}{June 03--05,
  2018}{Woodstock, NY}
\acmISBN{978-1-4503-XXXX-X/2018/06}

\begin{document}

\title{From User Behaviors to Satisfaction Judgments: Reasoning about User Satisfaction with AI Overviews}

\author{Xingzhu	Wang}
\email{wangxingzhu2022@ruc.edu.cn}
\affiliation{%
  \institution{Renmin University of China Gaoling School of Artificial Intelligence}
  \country{China}
}
\author{Yitong Xu}
\email{xuyitong@baidu.com}
\affiliation{%
  \institution{Baidu Inc.}
  \country{China}
}

\author{Yue	Li}
\email{liyue50@baidu.com}
\affiliation{%
  \institution{Baidu Inc.}
  \country{China}
}

\author{Daiting Shi}
\email{shidaiting01@baidu.com}
\affiliation{%
  \institution{Baidu Inc.}
  \country{China}
}

\author{Jingzhou He}
\email{hejingzhou@baidu.com}
\affiliation{%
  \institution{Baidu Inc.}
  \country{China}
}

\author{Jiaxin Mao}
\email{maojiaxin@gmail.com}
\affiliation{%
  \institution{Renmin University of China Gaoling School of Artificial Intelligence}
  \country{China}
}

\renewcommand{\shortauthors}{Trovato et al.}

\begin{abstract}
The growing presence of AI overviews in everyday search makes their effective evaluation essential. Response quality has been the primary focus of existing evaluation research, with only a small number of studies examining the user perspective. These evaluations rely on queries and response content, leaving the information conveyed by user behavior insufficiently utilized. This work investigates satisfaction prediction for AI overviews from a user-centered perspective and incorporates behavioral signals into the prediction process. Applying existing behavior-based satisfaction prediction methods from traditional information retrieval to this setting, however, presents limitations. We address these limitations with STaRS, an LLM-based approach designed specifically for predicting satisfaction with AI overviews. At its core is a structured reasoning procedure: individual behavioral signals are interpreted first, and their implications then inform the prediction of user satisfaction. Reinforcement learning further optimizes this procedure, while a reward shaping mechanism alleviates reward sparsity during training. Evaluation on two real-world datasets demonstrates that STaRS outperforms the comparison methods in satisfaction prediction, providing search providers with a practical means of accurately estimating users' satisfaction with AI overviews.
\end{abstract}


\received{20 February 2007}
\received[revised]{12 March 2009}
\received[accepted]{5 June 2009}

\maketitle
\vspace{-8pt}
\section{Introduction}

\begin{figure}[H]
\vspace{-9pt}
  \centering
 \includegraphics[width=\linewidth]{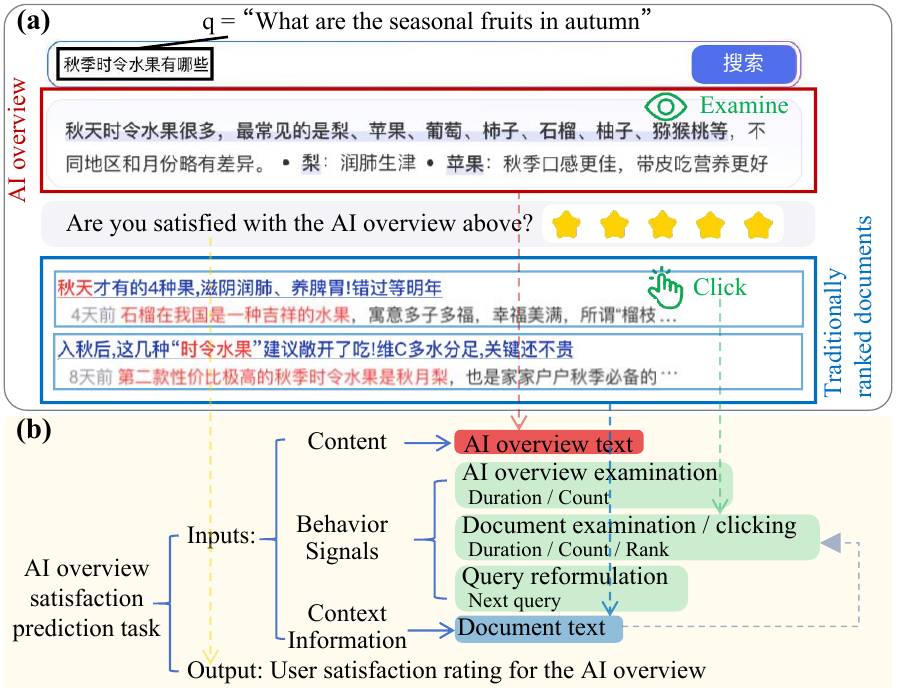}
 \vspace{-18pt}
\caption{
Illustration of (a) a SERP containing an AI overview (with a satisfaction rating card) and (b) inputs and output for our AI overview satisfaction prediction task.
}
  \label{fig:input}
\end{figure}

Generative Artificial Intelligence (GenAI) is increasingly being integrated into information access. In web search, one prominent application of GenAI is the AI overview, a generated summary presented above traditionally ranked results (as shown in Figure~\ref{fig:input}(a)) on the search engine results page (SERP) with the goal of directly addressing the user's information need. 
This hybrid retrieval setting, which jointly presents an AI overview alongside traditional ranked results, represents a new form of information access that differs from both traditional IR, which presents only ranked results, and ChatGPT-style conversational search, which presents only a generated response.
As AI overviews are increasingly adopted by major search engine companies in real-world search systems and become a more common part of users' everyday search experiences, their effective evaluation becomes increasingly important. 
\vspace{-12pt}
As the quality of AI overviews continues to improve, evaluating and improving users’ actual experiences with these overviews has not received commensurate attention. Existing research offers a wide range of system-centric methods for evaluating AI-generated responses, commonly applied to AI overviews to assess relevance, faithfulness, and completeness~\cite{saad-falcon-etal-2024-ares, ferrara2024ragtriad, es-etal-2024-RAGAs, ispas2025towards}.
Few studies, however, have explored user-centric evaluation~\cite{10.1145/3626772.3657849,wang-etal-2024-user}.
This gap limits opportunities to improve user experience and retention on AI overview ~\cite{dan2016measuring}, as system-centric quality criteria cannot fully capture the user perspective on AI overviews~\cite{mao2016does,10.1145/3077136.3080804}.
User satisfaction, in particular, is widely regarded as the gold standard for IR evaluation~\cite{zhang2020models}.
Yet, to the best of our knowledge, AI overview satisfaction prediction has not been investigated in prior work.

To address this gap, we move beyond system-centric response quality assessment toward a user-centric perspective and predict satisfaction with AI overviews. In addition, existing AI overview evaluation methods largely rely on queries and response content, overlooking informative user behavior~\cite{10.1145/2600428.2609607}, which may limit evaluation accuracy~\cite{10.1145/1718487.1718515,10.1145/3077136.3080804}.
Accordingly, we formulate the task of AI overview satisfaction prediction, which differs from existing AI overview evaluation tasks in two aspects:
(1) We directly predict satisfaction with AI overviews, rather than assessing response quality or other user-centric criteria.
(2) We incorporate user behavior as an important input alongside the query and response content. 

Although behavior-based satisfaction prediction methods have been widely studied in traditional IR~\cite{jiang2015understanding,10.1145/2766462.2767721, kim2014modeling,kiseleva2016predicting}, applying them to AI overviews remains challenging, as what a behavior signal implies about satisfaction in traditional IR may not transfer to AI overview satisfaction prediction. 
For example, when assessing document or query satisfaction, a long document dwell time ($\mathrm{dwell}>30\,\mathrm{s}$) suggests engagement and perceived document usefulness
\footnote{Interpretation refers to the user's state or perception of the AI overview inferred from a behavioral signal.}
~\cite{fox2005evaluating,10.1145/1718487.1718515} and is considered a positive signal
\footnote{Polarity denotes whether a behavioral signal is positive, negative, or neutral with respect to user satisfaction, based on its interpretation.}. For AI overviews, however, the same behavior may instead signal dissatisfaction if the document addresses information needs left unmet by the AI overview. It is also possible that $\mathrm{dwell}>30$ may indicate satisfaction if the overview sparks curiosity about a related topic beyond the scope of the user’s original information need, prompting the user to click a document for further exploration.
These examples show that a positive behavior signal in traditional IR may indicate either satisfaction or dissatisfaction with an AI overview.
As a result, this limits the direct applicability and effectiveness of existing behavior-based methods, such as logistic regression using document dwell time as a numerical feature~\cite{jiang2015understanding,10.1145/2766462.2767721}.
Accordingly, methods that reinterpret observed behavior for AI overview satisfaction prediction are needed.

Recent advances in LLM reasoning~\cite{deepseekai2025deepseekv32pushingfrontieropen} offer a promising way to address this challenge. By jointly considering the query, AI overview, and associated context information, LLMs with reasoning ability can help reinterpret observed behaviors, for example, distinguishing whether a click reflects an unmet information need or further exploration prompted by the overview, thereby enabling more accurate prediction. Accordingly, we formulate AI overview satisfaction prediction as a reasoning task and propose a novel LLM-based method. We propose \textbf{STaRS}, which consists of a \textbf{ST}ructured re\textbf{a}soning strategy and a \textbf{R}eward \textbf{S}haping mechanism for reinforcement learning (RL). Specifically, we design a structured reasoning strategy that guides LLMs to first reason over individual behavior signals at a finer granularity and then predict overall satisfaction score, yielding a more faithful estimation.
To further improve performance, we optimize the model through RL.
Furthermore, since rewards based solely on the final prediction are sparse and thus provide limited guidance for intermediate reasoning over behavior signals, we design a reward-shaping mechanism that introduces supervision on the satisfaction polarities of influential behavior signals, thereby enhancing the trained model's reasoning ability. 

In summary, we make three main contributions:

To the best of our knowledge, we are the first to investigate the task of predicting user satisfaction with AI overviews on SERPs.
To this end, we construct two new datasets from complementary sources: a laboratory user study and industrial search logs.

We propose STaRS. We design a novel structured reasoning strategy to reason over user behavior signals for more accurate satisfaction prediction. We develop a new reward shaping mechanism for this reasoning structure to address reward sparsity in RL.

Experiments on both datasets show that STaRS, tailored to AI overview satisfaction prediction, outperforms existing evaluation methods across three categories, suggesting potential value for providers of AI overview services.
\vspace{-6pt}
\section{Related Work}

\subsection{User Satisfaction}
Prior work has conceptualized search satisfaction as users’ perceived fulfillment of their information needs ~\cite{10.1145/3336191.3371799} and treated it 
as a gold standard for evaluating IR systems~\cite{dan2016measuring,  jiang2015understanding, zhang2020models}. Predicting satisfaction is an important task and helps monitor user experience and guide IR system improvements~\cite{dan2016measuring,  10.1145/2600428.2609607}. Prior work has explored predicting document~\cite{fox2005evaluating}, query~\cite{hassan2013beyond,mehrotra2017user,chen2017user}, and session satisfaction~\cite{jiang2015understanding, fox2005evaluating} using behavior signals from search logs.
In this work, we focus on predicting satisfaction with AI overviews.
\vspace{-8pt}
\subsection{Behavior-Based Satisfaction Prediction Methods in Traditional IR}
Prior work in traditional IR has widely studied interaction signals as implicit indicators of  satisfaction~
\cite{hassan2013beyond,mehrotra2017user,fox2005evaluating,chakraborty2014correlation,xu2009evaluating,10.1145/1718487.1718515,chen2017user,huang2012improving,lagun2014towards}, including a top-result click as a positive satisfaction signal~\cite{mehrotra2017user}, a similar query reformulation issued in a short time interval as an indicator of  dissatisfaction~\cite{hassan2013beyond}, and a long dwell time on a document as an indicator of satisfaction~\cite{fox2005evaluating}.
Building on these signals, numerous studies have hand-crafted numerical features and trained machine-learning models, including logistic regression~\cite{jiang2015understanding,10.1145/2766462.2767721}, tree-based models such as GBDT~\cite{kim2014modeling,kiseleva2016predicting}, and sequence models~\cite{10.1145/2736277.2741669}, to predict satisfaction. Prior studies have demonstrated that behavior signals can reflect satisfaction more closely than content quality measures across various search settings~\cite{10.1145/1718487.1718515,10.1145/3209978.3210059,10.1145/3077136.3080804}, and behavior-based satisfaction prediction has received considerable attention~\cite{10.1145/2600428.2609607}, motivating us to incorporate behavior signals for AI overview satisfaction prediction in this paper.

However, when the prediction target shifts to AI overview satisfaction, the interpretations of these behavior signals in traditional IR may not apply directly to AI overviews, and numerical behavior features alone may not adequately capture AI overview satisfaction. Therefore, in this paper, we develop STaRS, a new LLM-based method tailored to AI overview satisfaction prediction.

\vspace{-8pt}
\subsection{Evaluation Methods for AI Overviews}
\noindent\textbf{System-Centric Quality Evaluation Methods.}
There has been a surge of research interest in evaluating the quality of AI-generated responses from a system-centric perspective~\cite{ferrara2024ragtriad, saad-falcon-etal-2024-ares, es-etal-2024-RAGAs, ispas2025towards, 3737916.3738608}%
~and these methods are commonly applied to evaluate AI overviews. For example, TruLens assesses context relevance, groundedness, and answer relevance~\cite{ferrara2024ragtriad}, while RAGAs operationalizes similar dimensions through evaluation that does not require ground-truth answers~\cite{es-etal-2024-RAGAs}. 
More recently, Ispas et al.~\cite{ispas2025towards} investigate fine-grained statement-level evaluation of correctness and faithfulness.

\noindent\textbf{User-Centric Evaluation Methods.}
Only a few studies have explored user-centric evaluation of generated responses, extending system-centric assessment to consider how well these responses meet users' needs.
For example, Gienapp et al.~\cite{10.1145/3626772.3657849} propose a framework for evaluating the utility of generated responses across multiple user-relevant dimensions. 

Nevertheless, both types of evaluation methods discussed above rely solely on queries and AI-generated response content, overlooking the crucial role of user behavior signals in evaluation. Recently, G-CoS~\cite{10.1145/3805712.3809934} pioneers the incorporation of user behavior into satisfaction prediction for ChatGPT-like systems rather than AI overviews. Our work stands apart by directly predicting user satisfaction with AI overviews while incorporating user behavior signals.
\vspace{-8pt}
\subsection{Reward Shaping in RL}
Reward shaping is an effective approach to alleviating reward sparsity in RL~\cite{zheng2026comprehensive, yu2025reward, schoepp2025evolving}. Instead of relying solely on delayed outcome rewards, it augments the learning objective with intermediate feedback, providing denser reward signals that guide learning toward the desired outcome. 
Recently, a key motivation for its application to LLM reasoning is to improve intermediate reasoning quality.
For example, Search-P1~\cite{xia2026search} introduces path-centric rewards, providing learning signals from intermediate reasoning rather than relying solely on final outcomes. 
MENTOR~\cite{choi2025mentor} employs flexible teacher-guided rewards to provide process-level supervision for tool-use reasoning. 
Inspired by them, we also employ reward shaping to improve the quality of intermediate reasoning in STaRS.
\vspace{-8pt}
\section{Data}
To study AI overview satisfaction prediction, we collect real users’ interaction logs and their explicit satisfaction scores, in addition to the queries, AI overviews used for AI overview quality evaluation. Therefore, we construct two datasets from two distinct data sources: a user study designed for this task and an industrial search engine company with real users. Their statistics are shown in Table~\ref{tab:dataset_statistics}.  
\vspace{-8pt}
\subsection{Data Collection}
 \subsubsection{UserStudy Dataset}
The dataset contains 554 query records from 243 task sessions involving 31 participants. Each query record contains the query, the corresponding AI overview, traditional ranked search results, user behavior, and a satisfaction score reported by the user who performed the search.

\textbf{Experimental Procedure.}
The dataset was collected through a controlled user study: 
We first used an experimental search engine system following Liu et al.~\cite{10.1145/3292500.3330981} 
and modified its interface to present an AI overview alongside traditional ranked search results, as shown in Figure~\ref{fig:input}(a). We obtained both the AI overview and the ranked results through an API provided by a major commercial search engine and removed advertisements before display.
We then recruited participants to perform search tasks drawn from a set of 10 tasks of varying difficulty, with most tasks selected from the information needs in the TREC 2002--2004 collection~\cite{bailey2015information}. During the experiment, we instructed participants to interact with the system as they normally would with a search engine with AI overviews. 

\textbf{Data Collection Procedure.}
(1) User interaction logs: A JavaScript  browser plugin recorded users' actual interaction behavior while they engaged with the search system.
(2) First-party satisfaction scores with the AI overview: After each query, participants rated their satisfaction with the AI overview on a five-point scale.
\vspace{-2pt}
\subsubsection{SearchLog26 Dataset}
We constructed the SearchLog26 dataset using search logs from an anonymous company’s search engine with AI overviews.
The SearchLog26 dataset comprises 6{,}589 query-level search logs from 5.4k users. Each query record includes the same fields as those in the UserStudy dataset.

\textbf{Data Collection Procedure.}
(1) User interaction logs: The client recorded viewport, click, and other interaction events, along with their timestamps. (2) First-party satisfaction scores with the AI overview: Following prior work~\cite{white2012leaving,10.1145/2983323.2983829}, we displayed a pop-up questionnaire (``Feedback about your satisfaction with AI overview'') under the AI overview component as shown in Figure~\ref{fig:input}(a), asking users to rate their satisfaction with it on a five-point scale. Users' explicit satisfaction scores are used as ground-truth scores. Furthermore, for users who reported (dis)satisfaction scores, the questionnaire additionally asked them to specify the reasons for their scores, as shown in Figure~\ref{fig:app}. Users' explicit reasons are used as ground-truth reasons. 

\vspace{-10pt}
\begin{table}[H]
\centering
\caption{Dataset statistics. ``\#'' denotes counts,
and ``/Qry'' denotes averages per query (equivalently,
per AI overview).
``\#Beha-Signal'' denotes the number of behavior signals.}

\label{tab:dataset_statistics}
\setlength{\tabcolsep}{3pt}
\vspace{-10pt}
\resizebox{\columnwidth}{!}{%
\begin{tabular}{@{}ccccccc@{}}
\toprule
Datasets
& Source
& \shortstack{Satisfaction Scale}
& \#User
& \#Qry
& \#Beha-Signal
& \shortstack{\#Beha-Signal/Qry} \\
\midrule
SearchLog26
& Company
& 5-point
& 5.4K
& 6,589
& 26K
& 3.98 \\
UserStudy
& Laboratory
& 5-point
& 31
& 554
& 2,268
& 4.09  \\
\bottomrule
\end{tabular}%
}
\vspace{-12pt}
\end{table}
\vspace{-8pt}
\subsection{Behavior Signal Construction from Interaction Logs}
\label{sec:behavior_signal}
We construct four types of behavior signals from the collected interaction logs: \textbf{AI overview examination}, \textbf{document examination}, \textbf{document clicking}, and \textbf{query reformulation} (see Figure~\ref{fig:input}(b)), as prior studies have shown that these behaviors are associated with user satisfaction~\cite{fox2005evaluating,lagun2014towards,hassan2013beyond}.
Specifically, behavior signals are constructed and obtained from interaction logs as follows: (1) AI overview examination is characterized by duration, count, and rank, where viewport duration and viewport count are used for the SearchLog26 dataset, while exposure duration is used for the UserStudy dataset. (2) Document examination is characterized in the same way as AI overview examination. (3) Document clicking is characterized by click duration, click count, and document rank. (4) Query reformulation is represented by the subsequent query. 
\vspace{-6pt}

\section{Method}
Our goal is to predict AI overview satisfaction by incorporating user behavior. We first formulate the task and then detail STaRS.
\vspace{-8pt}
\subsection{Problem Formulation}
\textbf{Task Input.} 
For a query $q$, let $a$ denote the generated AI overview, $B=\{b_i\}_{i=1}^{N}$ denote the behavior signals defined in Section~\ref{sec:behavior_signal}, and $C=\{c_1,\ldots,c_N\}$ denote the corresponding contexts in which these behaviors occur. $N$ is the number of behavior signals\footnote{The number of behavior signals varies across samples; for simplicity, we use $N$ to denote this number for each sample $x$.}. Note that $B$ can be incorporated to predict AI overview satisfaction because prior studies in traditional web search have shown that users' perceived satisfaction, although not directly observable, is reflected in their subsequent search behavior signals $B$ ~\cite{awadallah2013personalized, hassan2013beyond}. Besides, we incorporate $C$ because behavior signals can be ambiguous in isolation, and the context information associated with these behavior signals helps them become clearer satisfaction indicators. Specifically, in this work, each $c_i$ contains the text of the corresponding exposed or clicked document from the traditional ranked results; $c_i=\varnothing$ for AI overview examination and query reformulation.

\textbf{Task Output.} The output is the user's satisfaction with the AI overview, denoted as \(\mathrm{Sat}\)\footnote{For brevity, we denote satisfaction by $y$ throughout the remainder of this paper.}, where
$
\mathrm{Sat} \in \{1,2,\ldots,L\},
$
and  \(L\) is the number of satisfaction levels.

Formally, our task aims to learn a function:
\begin{equation}
\widehat{\mathrm{Sat}} = f(q,a,B,C).
\end{equation}
Our task differs from the three existing tasks:
\textbf{AI overview system-centric quality evaluation}~\cite{ferrara2024ragtriad,saad-falcon-etal-2024-ares,es-etal-2024-RAGAs},
which focuses on $\widehat{\mathrm{Qua}} = f(q,a)$,
\textbf{AI overview user-centric evaluation}~\cite{10.1145/3626772.3657849,wang-etal-2024-user},
which focuses on $\widehat{\mathrm{Uce}} = f(q,a)$, where $\mathrm{Uce}$ denotes the user-centric evaluation score, and
\textbf{behavior-based satisfaction prediction in traditional IR}~\cite{10.1145/1718487.1718515,jiang2015understanding,mehrotra2017user},
which predicts 
$\widehat{\mathrm{Sat}}_{\mathrm{search}} = f(B)$.
\vspace{-8pt}
\subsection{Overview}
\begin{figure*}[h]
    \centering
    \includegraphics[width=\textwidth]{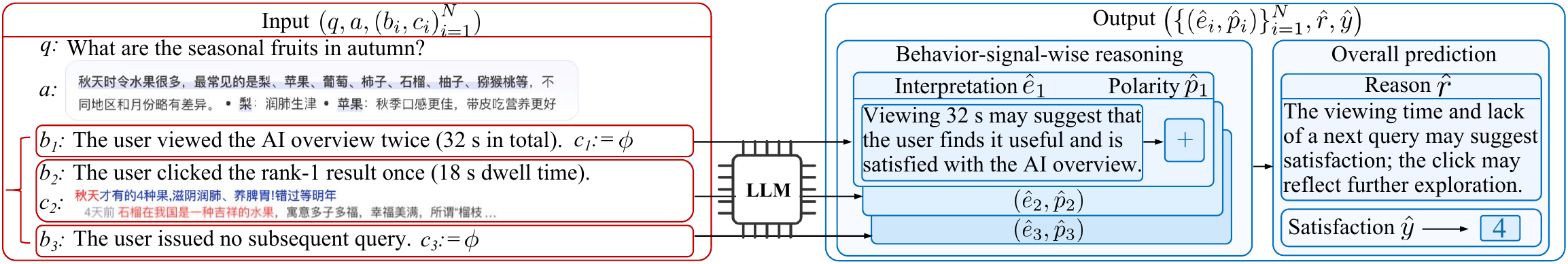}
    \vspace{-22pt}
\caption{\textbf{The proposed structured reasoning strategy.}
Given the query $q$, AI overview $a$, behavior signals $B$, and context information $C$ (red), the LLM is prompted to generate a structured reasoning trace (blue).}
\label{fig:reasoning-paradigm}
   \vspace{-10pt}
\end{figure*}

\begin{figure}[h]
    \centering
    \includegraphics[width=\linewidth]{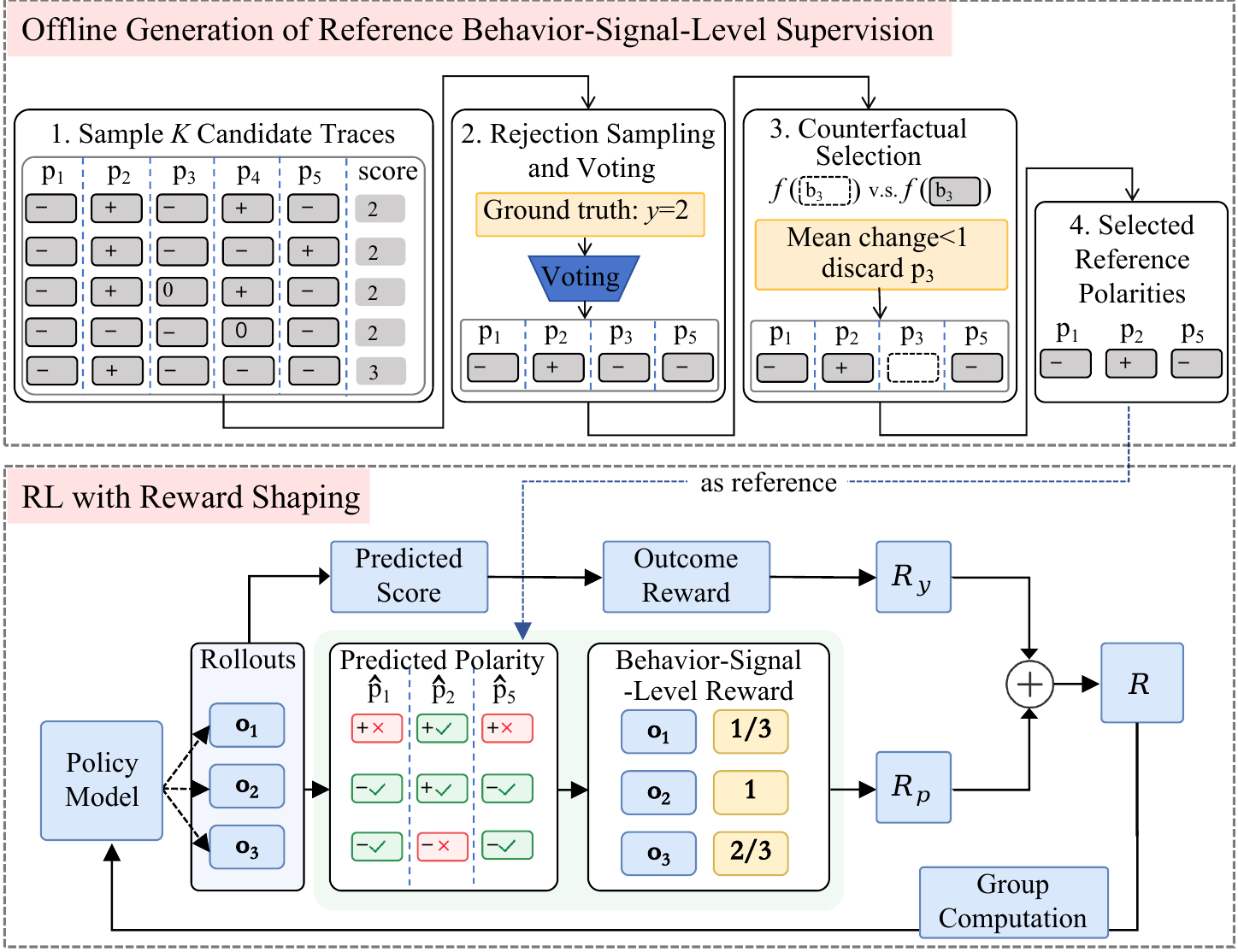}
       \vspace{-14pt}
    \caption{Illustration of the proposed reward shaping mechanism. \textbf{Top:} Offline generation of behavior-signal-level supervision. \textbf{Bottom:} RL with reward shaping. Gray indicates offline processing, blue denotes components of the naive GRPO framework, and the green-shaded region highlights our added behavior-signal-level reward computation.}
    \label{fig:reward_shaping}
       \vspace{-18pt}
\end{figure}

STaRS consists of two parts, namely a structured reasoning strategy and RL with a
reward shaping mechanism. Section ~\ref{sec:reasoning_paradigm} and Figure~\ref{fig:reasoning-paradigm} introduce our structured reasoning strategy for predicting AI overview satisfaction.
Section 
~\ref{sec:RL-reward-shaping} 
and Figure~\ref{fig:reward_shaping}
introduce our RL training with a reward shaping mechanism.
\vspace{-8pt}
\subsection{Structured Reasoning Strategy}
\label{sec:reasoning_paradigm}
Predicting satisfaction through a general instruction---such as ``Read the user's entire search process and infer their satisfaction with the AI overview''---encourages an LLM-based predictor to assess satisfaction holistically, which may limit prediction accuracy.
Wang et al.~\cite{10.1145/2600428.2609607} have
demonstrated that finer-grained action-level satisfaction modeling
can yield more accurate overall satisfaction predictions. Motivated by them, we explicitly  interpret what each behavior signal implies about satisfaction before predicting an overall satisfaction score.

As shown in Figure~\ref{fig:reasoning-paradigm}, for an instance $x=(q,a,B,C)$, we first present the input to LLMs in a structured format:
$( q,a,\{(b_i,c_i)\}_{i=1}^{N})$, in which each behavior signal $b_i$ constructed
as described in Section~\ref{sec:behavior_signal} is expressed in natural language to form the input. LLMs are then prompted to generate a structured reasoning trace in two steps:
(1) deriving an interpretation $\hat e_i$ of each behavior signal $b_i$ using the query $q$, AI overview $a$, and context information $c_i$, and assigning a satisfaction polarity $\hat p_i$ based on this interpretation;
(2) synthesizing these interpretations and polarities into an overall reason $\hat r$ and then predicting the overall satisfaction score $\hat y$. The resulting structured reasoning trace is:
\begin{equation}
    \hat T=
    \left(
        \{(\hat e_i,\hat p_i)\}_{i=1}^{N},
        \hat r,\hat y
    \right).
    \label{eq:reasoning-trace}
\end{equation}
Here, $\hat e_i$ explains the satisfaction implication of behavior signal $b_i$, and $\hat p_i\in \{+,-,0\}$ indicates the satisfaction polarity reflected by the behavior signal: satisfaction with the AI overview, dissatisfaction with it, or no clear direction, respectively. $\hat y\in\{1,\ldots,L\}$. The detailed full prompt and output format are provided in Table~\ref{tab:structured_reasoning_prompt}.

\vspace{-8pt}
\subsection{
RL with a Reward Shaping Mechanism
}
\label{sec:RL-reward-shaping}
\subsubsection{Optimizing Satisfaction Prediction Model with RL}
\label{sec:rl-optimization}

To effectively optimize this structured reasoning process, we further train LLMs for AI overview satisfaction prediction through RL from verifiable rewards (RLVR), using Group Relative Policy Optimization (GRPO)~\cite{shao2024deepseekmathpushinglimitsmathematical} to maximize the following objective:
\begin{equation}
\begin{aligned}
\mathcal{J}_{\mathrm{GRPO}}(\theta)
=
&
\mathbb{E}_{x\sim\mathcal{D},\,\{o_i\}_{i=1}^{G}\sim\pi_{\theta_{\mathrm{old}}}(\cdot\mid x)}
\Bigg[
\frac{1}{G}
\sum_{i=1}^{G}
\frac{1}{|o_i|}
\sum_{t=1}^{|o_i|} \\[-0.8ex]
&\min\Big(
r_{i,t}(\theta)\hat{A}_i,
\operatorname{clip}\left(
r_{i,t}(\theta),1-\epsilon,1+\epsilon
\right)\hat{A}_i
\Big)
\Bigg].
\end{aligned}
\end{equation}
\begin{equation}
r_{i,t}(\theta)
=
\frac{
\pi_{\theta}\left(o_{i,t}\mid x,o_{i,<t}\right)
}{
\pi_{\theta_{\mathrm{old}}}\left(o_{i,t}\mid x,o_{i,<t}\right)
}.
\end{equation}

\begin{equation}
\hat{A}_i
=
\frac{
R_i-\operatorname{mean}\left(\{R_j\}_{j=1}^{G}\right)
}{
\operatorname{std}\left(\{R_j\}_{j=1}^{G}\right)
},
\end{equation}

where $x=(q,a,B,C)$ denotes a complete input instance sampled from the training dataset $\mathcal{D}$, $o_i=(o_{i,1},\ldots,o_{i,|o_i|})$ denotes the $i$-th
generated response which is viewed as a generation trajectory, $r_{i,t}(\theta)$ denotes the importance sampling ratio for the $t$-th token in the $i$-th generated response, and $\hat{A}_i$ denotes the normalized advantage of the $i$-th generated response.
\subsubsection{Reference Behavior-Signal-Level Supervision Generation}
\label{sec:reward-shaping}
In a naive setup using only an outcome reward, GRPO provides sparse reward based solely on final prediction correctness,
ignoring intermediate reasoning quality~\cite{xia2026search}. For example, a rollout may correctly interpret AI overview examination, clicks, and query reformulations yet receive a low reward for predicting an incorrect satisfaction level. Conversely, a rollout that misinterprets these behavior signals may receive a higher reward if it happens to predict the correct satisfaction level. Such misleading rewards can reinforce flawed reasoning, impairing both the model's ability to interpret behavioral signals and to predict satisfaction. To alleviate it, we introduce behavior-signal-level reward shaping to provide finer-grained rewards. The reference supervision is generated and filtered offline using a high-capability LLM through the following three stages. 

\textbf{In the first stage}, to obtain behavior-signal-level supervision for constructing the behavior-signal-level reward, we begin by generating candidate reference traces. For each training
instance $x$, we use the high-capability LLM as the reference model to sample $K$ candidate traces following
the same structured reasoning strategy:
\begin{equation}
\begin{aligned}
    \mathcal{T}_{\mathrm{ref}}
    &= \{T_{\mathrm{ref},k}\}_{k=1}^{K}, \\
    T_{\mathrm{ref},k}
    &=
    \left(
        \{(e_{\mathrm{ref},k,i},p_{\mathrm{ref},k,i})\}_{i=1}^{N},
        r_{\mathrm{ref},k},y_{\mathrm{ref},k}
    \right),
\end{aligned}
\label{eq:reference-candidates}
\end{equation}
where $\mathcal{T}_{\mathrm{ref}}$ denotes the collection of $K$ candidate reference traces generated for $x$, $k=1,\ldots,K$.
The components of this trace are defined as in Eq.~\ref{eq:reasoning-trace}. These candidate traces will be subsequently used to construct final behavior-signal-level reference supervision.

\textbf{ In the second stage}, to obtain reliable reference supervision from candidate traces, we use rejection sampling and majority voting inspired by Xia et al.~\cite{xia2026search}. (1) We retain only traces whose predicted satisfaction scores match the ground truth. Formally, the set of retained traces is defined as follows:
\begin{equation}
\mathcal{T}_{\mathrm{ref}}^{+}
=
\left\{
T_{\mathrm{ref},k}
\;\middle|\;
1\leq k\leq K,\; y_{\mathrm{ref},k}=y
\right\},
\label{eq:reference-filtering}
\end{equation}
where $\mathcal{T}_{\mathrm{ref}}^{+}$ contains the candidate traces whose predicted satisfaction scores match the ground-truth score $y$.
Let \(K'=|\mathcal{T}_{\mathrm{ref}}^{+}|\) denote the number of retained traces. For the voting step, we re-index these traces by \(k=1,\ldots,K'\). (2) To obtain the reference polarities for constructing the final behavior-signal-level reference supervision, we perform a voting procedure as follows. We first count the occurrences of each polarity for each behavior signal across the retained traces in $\mathcal{T}_{\mathrm{ref}}^{+}$. We then retain a reference polarity only if more than half of the retained traces assign that polarity to the behavior signal:
\begin{equation}
\begin{aligned}
I_{\mathrm{vote}}
&=
\left\{
i\in\{1,\ldots,N\}
\;\middle|\;
\frac{1}{K'}
\sum_{k=1}^{K'}
\mathbf{1}
[p_{\mathrm{ref},k,i}=p_{\mathrm{ref},i}]
>\frac{1}{2}
\right\},\\
P_{\mathrm{ref}}^{\mathrm{vote}}
&=
(p_{\mathrm{ref},i})_{i\in I_{\mathrm{vote}}}.
\end{aligned}
\label{eq:reference-voting}
\end{equation}
Here, $i=1,\ldots,N$ indexes the behavior signals, $p_{\mathrm{ref},i}\in\{+,-,0\}$, $p_{\mathrm{ref},k,i}$ is the polarity assigned by the $k$-th retained trace, and $\mathbf{1}[\cdot]$ is the indicator function. $I_{\mathrm{vote}}$ is the set of majority-voted signal indices, and $P_{\mathrm{ref}}^{\mathrm{vote}}$ is the corresponding indexed polarity collection.

 \textbf{In the third stage}, to retain reference polarities associated with influential behavior signals as the final reference supervision, we perform counterfactual signal selection. Because even consensus reference polarities may correspond to peripheral or redundant behavior signals; rewarding all of them equally may dilute supervision on influential signals. We assess the importance of each behavior signal by comparing satisfaction scores re-estimated before and after counterfactually removing the behavior signal.

Specifically, let $f_{\mathrm{ref}}$ denote re-estimation by the high-capability LLM, which uses the prompt in Table~\ref{tab:structured_reasoning_prompt} augmented with reference polarities $P_{\mathrm{ref}}^{\mathrm{vote}}$ or $P_{\mathrm{ref}}^{\mathrm{vote}}\setminus\{p_{\mathrm{ref},j}\}$, keeping the provided polarities fixed, inferring the rest, and returning the re-estimated satisfaction score. For each behavior signal \(b_j\) with \(j\in I_{\mathrm{vote}}\), we use $f_{\mathrm{ref}}$ to re-estimate satisfaction before and after counterfactual removal, repeating each prediction \(Z\) times to reduce prediction randomness. We then average the scores over the \(Z\) repetitions for each condition and compute the absolute difference between the two means:
\begin{equation}
\begin{aligned}
y_{\mathrm{ref},z}^{\mathrm{vote}}
&=
f_{\mathrm{ref}}\left(
q,a,B',C',
P_{\mathrm{ref}}^{\mathrm{vote}}
\right),
\\
y_{\mathrm{ref},z}^{-j}
&=
y_{\mathrm{ref},z}^{\operatorname{do}(b_j=\varnothing)}
\\
&=
f_{\mathrm{ref}}\Bigl(q,a,B'\setminus\{b_j\},C'\setminus\{c_j\},P_{\mathrm{ref}}^{\mathrm{vote}}\setminus\{p_{\mathrm{ref},j}\}\Bigr),
\\
\Delta_{\mathrm{ref},j}
&=
\left|
\frac{1}{Z}\sum_{z=1}^{Z}y_{\mathrm{ref},z}^{-j}
-
\frac{1}{Z}\sum_{z=1}^{Z}y_{\mathrm{ref},z}^{\mathrm{vote}}
\right|,
\quad j\in I_{\mathrm{vote}},
\end{aligned}
\label{eq:counterfactual-selection}
\end{equation}
where $I_{\mathrm{vote}}$ contains the indices of the signals kept after voting, and $B'$ and $C'$ contain these behavior signals and their context information, respectively.
The index $z=1,\ldots,Z$ indexes repeated predictions.
The operator $\operatorname{do}(b_j=\varnothing)$ denotes the counterfactual removal of $b_j$ from the reference model's input; its corresponding context information $c_j$ and reference polarity $p_{\mathrm{ref},j}$ are removed accordingly.
The scores $y_{\mathrm{ref},z}^{\mathrm{vote}}$ and $y_{\mathrm{ref},z}^{-j}$ are predicted before and after removal, respectively, and $\Delta_{\mathrm{ref},j}$ is the absolute difference between their means.

Finally, we retain only non-neutral reference polarities associated with behavior signals satisfying $\Delta_{\mathrm{ref},j}\geq1$ to form the final reference supervision:
\begin{equation}
\begin{aligned}
I_{\mathrm{ref}}
&=
\left\{
j\in I_{\mathrm{vote}}
\;\middle|\;
\Delta_{\mathrm{ref},j}\geq1,\;
p_{\mathrm{ref},j}\neq0
\right\},\\
P_{\mathrm{ref}}
&=
(p_{\mathrm{ref},j})_{j\in I_{\mathrm{ref}}},
\end{aligned}
\label{eq:reference-selection}
\end{equation}
where the index sets satisfy
$I_{\mathrm{ref}}\subseteq I_{\mathrm{vote}}
\subseteq\{1,\ldots,N\}$ and \textbf{$P_{\mathrm{ref}}$ is the final indexed
collection of reference supervision},
preserving the original signal indices.
Note that if no candidate trace matches the ground-truth score in the first stage, we skip voting and counterfactual signal selection and set $P_{\mathrm{ref}}$ to $\varnothing$ for that instance.

In essence, by our counterfactual selection, we identify behavior signals whose misinterpretation may distort the final satisfaction prediction, and by rewarding agreement with the reference polarities for these influential signals, we thereby achieve more accurate final satisfaction predictions. Next, we introduce our reward
design.

\subsubsection{Reward for RL}
\label{sec:rl-rewards}
The overall reward for RL training combines the format, outcome, and behavior-signal-level rewards.
\begin{itemize}
\item  \textbf{Format Reward}. The format reward $R_f$ measures whether the output is valid. An output is valid if it can be successfully parsed into its constituent fields without syntax errors.
\begin{equation}
R_f(\hat T)=
\begin{cases}
1, & \text{if } \hat T \text{ is valid},\\
0, & \text{otherwise}.
\end{cases}
\label{eq:format-reward}
\end{equation}
\item \textbf{Behavior-Signal-Level Reward.}
For a valid trace $\hat T$, $R_p$ is the proportion of retained reference polarities matched by the predicted polarities for the same behavior signals:

\begin{equation}
R_p(\hat T,P_{\mathrm{ref}})
=
\begin{cases}
\displaystyle
\frac{1}{|I_{\mathrm{ref}}|}
\sum_{i\in I_{\mathrm{ref}}}
\mathbf{1}[\hat p_i=p_{\mathrm{ref},i}],
& |I_{\mathrm{ref}}|>0,\\[8pt]
0,
& |I_{\mathrm{ref}}|=0,
\end{cases}
\label{eq:process-reward}
\end{equation}
where $I_{\mathrm{ref}}$ and $P_{\mathrm{ref}}$ are obtained from Eq.~\ref{eq:reference-selection}.

\item  \textbf{Outcome Reward.} The outcome reward encourages accurate satisfaction prediction
by converting the absolute prediction error into a normalized score:
\begin{equation}
    R_y(\hat y,y)=1-\frac{|\hat y-y|}{L-1}.
    \label{eq:outcome-reward}
\end{equation}
Since $\hat y,y\in\{1,\ldots,L\}$ with $L>1$, the maximum
possible absolute error is $L-1$. Dividing by $L-1$ normalizes
the error to $[0,1]$, and subtracting it from $1$ assigns higher
rewards to more accurate predictions.
\end{itemize}

We combine the three rewards for each generated trace.
\begin{equation}
    R(\hat T)=
    \begin{cases}
        \lambda_\alpha R_y(\hat y,y)
        +\lambda_\beta R_p(\hat T,P_{\mathrm{ref}}),
        & R_f(\hat T)=1,\\
        -1, & \text{otherwise}.
    \end{cases}
    \label{eq:total-reward}
\end{equation}
Here, $\lambda_\alpha,\lambda_\beta\geq0$ weight the outcome and behavior-signal-level
rewards, respectively. The label $y$ and reference
supervision $P_{\mathrm{ref}}$ are fixed for all rollouts of each instance. The policy model receives the full input \(x\) and predicts polarities for all \(N\) behavior signals; only signals indexed by \(I_{\mathrm{ref}}\) contribute to \(R_p\).
Note that since polarities are derived from interpretations, our reward shaping encourages better interpretations of behavior signals by rewarding agreement with reference polarities, thereby improving intermediate reasoning quality. This effect is supported by the experimental results in Appendix~\ref{sec:reward_shaping_reasoning_quality}.
\vspace{-8pt}
\section{Experiments}
\subsection{Baselines}
To evaluate the effectiveness of STaRS, we compare it with the following three types of baselines:
(1) \textbf{AI Overview Quality Evaluation Methods.}
These methods evaluate the quality of AI overviews using the query and the AI overview content, without incorporating user behavior as input.
\textbf{ARES}~\cite{saad-falcon-etal-2024-ares} evaluates context relevance, answer faithfulness, and answer relevance.
\textbf{TruLens}~\cite{ferrara2024ragtriad} measures context relevance, groundedness, and answer relevance.
\textbf{RAGAs}~\cite{es-etal-2024-RAGAs} uses LLM-based metrics that do not require ground-truth answers to assess faithfulness, answer relevance, and context relevance.
\textbf{OpenEval} denotes the evaluator proposed by Ispas et al.~\cite{ispas2025towards}, which uses fine-grained, statement-level judgments to evaluate correctness and faithfulness.
\vspace{-10pt}
\begin{table}[H]
\centering
\caption{Overall performance on the UserStudy dataset.}
\vspace{-10pt}
\resizebox{\linewidth}{!}{
\begin{tabular}{lllllll}
\hline
 & Precision & Recall & F1-Score & P-$r$ & S-$\rho$ & MAE \\
\hline
\multicolumn{7}{l}{\textbf{AI Overview Quality Evaluation Methods}} \\
ARES (context relevance)~\cite{saad-falcon-etal-2024-ares} & 0.2046$^{**}$ & 0.2181$^{**}$ & 0.2003$^{**}$ & 0.2261$^{**}$ & 0.2323$^{**}$ & 0.9693$^{**}$ \\
ARES (answer faithfulness)~\cite{saad-falcon-etal-2024-ares} & 0.2142$^{**}$ & 0.2240$^{**}$ & 0.1846$^{**}$ & 0.0374$^{**}$ & 0.0733$^{**}$ & 1.1787$^{**}$ \\
ARES (answer relevance)~\cite{saad-falcon-etal-2024-ares} & \textbf{0.4112} & 0.2470 & 0.2226 & 0.2921$^{**}$ & 0.2729$^{**}$ & 1.0000$^{**}$ \\
TruLens (context relevance)~\cite{ferrara2024ragtriad} & 0.2108$^{**}$ & 0.2203$^{**}$ & 0.2005$^{**}$ & 0.0961$^{**}$ & 0.1199$^{**}$ & 1.0648$^{**}$ \\
TruLens (groundedness)~\cite{ferrara2024ragtriad} & 0.2386$^{*}$ & 0.2239$^{**}$ & 0.1878$^{**}$ & 0.1275$^{**}$ & 0.1502$^{**}$ & 1.0848$^{**}$ \\
TruLens (answer relevance)~\cite{ferrara2024ragtriad} & 0.1178$^{**}$ & 0.2031$^{**}$ & 0.1137$^{**}$ & 0.2311$^{**}$ & 0.2671$^{**}$ & 1.1614$^{**}$ \\
RAGAs (faithfulness)~\cite{es-etal-2024-RAGAs} & 0.3700 & 0.2117$^{**}$ & 0.1739$^{**}$ & 0.0838$^{**}$ & 0.0857$^{**}$ & 1.1697$^{**}$ \\
RAGAs (answer relevance)~\cite{es-etal-2024-RAGAs} & 0.2502$^{*}$ & 0.2057$^{**}$ & 0.1410$^{**}$ & 0.1232$^{**}$ & 0.1079$^{**}$ & 1.1765$^{**}$ \\
RAGAs (context relevance)~\cite{es-etal-2024-RAGAs} & 0.1943$^{**}$ & 0.2024$^{**}$ & 0.1732$^{**}$ & 0.0300$^{**}$ & 0.0222$^{**}$ & 1.1697$^{**}$ \\
OpenEval (faithfulness)~\cite{ispas2025towards} & 0.2258$^{**}$ & 0.2183$^{**}$ & 0.1948$^{**}$ & 0.1447$^{**}$ & 0.1501$^{**}$ & 1.0247$^{**}$ \\

\hline
\multicolumn{7}{l}{\textbf{AI Overview User-Centric Evaluation Methods}} \\
GAH (coherence)~\cite{10.1145/3626772.3657849} & 0.1924$^{**}$ & 0.2214$^{**}$ & 0.1630$^{**}$ & 0.1554$^{**}$ & 0.1503$^{**}$ & 1.1462$^{**}$ \\
GAH (coverage)~\cite{10.1145/3626772.3657849} & 0.2269$^{**}$ & 0.2224$^{**}$ & 0.2152$^{*}$ & 0.2343$^{**}$ & 0.2318$^{**}$ & 0.9910$^{**}$ \\
GAH (consistency)~\cite{10.1145/3626772.3657849} & 0.2322$^{*}$ & 0.2030$^{**}$ & 0.1339$^{**}$ & 0.0330$^{**}$ & 0.0175$^{**}$ & 1.2310$^{**}$ \\
GAH (correctness)~\cite{10.1145/3626772.3657849} & 0.2991 & 0.2165$^{**}$ & 0.1530$^{**}$ & 0.0723$^{**}$ & 0.0571$^{**}$ & 1.2040$^{**}$ \\
GAH (clarity)~\cite{10.1145/3626772.3657849} & \underline{0.3884} & 0.2123$^{**}$ & 0.1237$^{**}$ & 0.0492$^{**}$ & 0.0600$^{**}$ & 1.2509$^{**}$ \\
GAH (overall)~\cite{10.1145/3626772.3657849} & 0.1652$^{**}$ & 0.2179$^{**}$ & 0.1597$^{**}$ & 0.1770$^{**}$ & 0.1896$^{**}$ & 1.1285$^{**}$ \\
URS~\cite{wang-etal-2024-user} & 0.2669$^{**}$ & 0.2401$^{*}$ & 0.2295 & 0.3259$^{*}$ & 0.3306$^{*}$ & 0.8917 \\
Direct Prompting~\cite{10.1145/3626772.3657707}  & 0.2459$^{**}$ & 0.2379$^{*}$ & 0.2089$^{**}$ & 0.2264$^{**}$ & 0.2375$^{**}$ & 0.9495$^{*}$ \\
Chain-of-Thought~\cite{3600270.3602070} & 0.2376$^{**}$ & 0.2529 & 0.2314 & 0.2856$^{**}$ & 0.2992$^{**}$ & 0.9134 \\
Rubric-Based Evaluation~\cite{gunjal2025rubricsrewardsreinforcementlearning} & 0.2468$^{**}$ & 0.2408$^{*}$ & 0.2256$^{*}$ & 0.2985$^{**}$ & 0.3087$^{**}$ & 0.9043 \\
G-Eval-Sat~\cite{liu-etal-2023-g} & 0.2568$^{**}$ & 0.2509 & 0.2439 & 0.3183$^{*}$ & 0.3187$^{**}$ & 0.9097 \\
\hline
\multicolumn{7}{l}{\textbf{User-Centric Satisfaction Prediction Methods Incorporating User Behaviors}} \\
G-CoS~\cite{10.1145/3805712.3809934} & 0.2986 & 0.2370$^{*}$ & 0.2162$^{*}$ & 0.3260$^{*}$ & 0.3407$^{*}$ & 0.9043 \\
Click-Reformulation Heuristic & 0.2648$^{**}$ & 0.2625 & \underline{0.2573} & 0.2281$^{**}$ & 0.2532$^{**}$ & 1.0812$^{**}$ \\
Dwell-Time Heuristic & 0.2092$^{**}$ & 0.2260$^{*}$ & 0.2038$^{**}$ & 0.1972$^{**}$ & 0.2027$^{**}$ & 1.0392$^{**}$ \\
Ridge & 0.2477$^{**}$ & 0.2341$^{*}$ & 0.1784$^{**}$ & 0.3101$^{*}$ & 0.3226$^{*}$ & 0.8978 \\
LSTM & 0.3433 & 0.2590 & 0.2119$^{*}$ & 0.3172$^{*}$ & 0.3352$^{*}$ & 0.8903 \\
Random Forest & 0.1291$^{**}$ & 0.2141$^{**}$ & 0.1306$^{**}$ & 0.3053$^{*}$ & \underline{0.3466} & 0.9047 \\
MLP & 0.2032$^{**}$ & 0.2247$^{**}$ & 0.1764$^{**}$ & 0.2451$^{**}$ & 0.2783$^{**}$ & 0.9236$^{*}$ \\
Transformer & 0.3346 & \underline{0.2725} & 0.2505 & \underline{0.3307} & 0.3342$^{*}$ & 0.8852 \\
GBDT & 0.3442 & 0.2562 & 0.2217$^{*}$ & 0.2929$^{*}$ & 0.2968$^{**}$ & \underline{0.8774} \\
\hline
STaRS & 0.3833 & \textbf{0.2783} & \textbf{0.2650} & \textbf{0.4051} & \textbf{0.4157} & \textbf{0.8646} \\
\hline
\end{tabular}
}
\par\vspace{2pt}
\begin{minipage}{\linewidth}
\footnotesize
$^{*}p < 0.05$, $^{**}p < 0.01$.
Stars on baseline entries indicate that STaRS is significantly better on that metric (sample-level paired bootstrap test, \(B = 10{,}000\) resamples).
\end{minipage}
\vspace{-14pt}
\label{tab:overall_userstudy}
\end{table}
(2) \textbf{AI Overview User-Centric Evaluation Methods.}
These existing methods, although adopting a user-centric perspective, still rely on the query and AI overview content for evaluation while overlooking user behavior.
\textbf{GAH} denotes Gienapp et al.'s utility evaluation framework~\cite{10.1145/3626772.3657849}, which scores responses across five dimensions.
\textbf{URS} adapts the URS benchmark's evaluation procedure~\cite{wang-etal-2024-user} to assess whether responses satisfy intent-specific user needs.
Although user-centric, the baselines described above do not directly predict user satisfaction; we therefore also include the following satisfaction prediction baselines:
\textbf{Direct Prompting} adapts Thomas et al.'s LLM-based evaluation~\cite{10.1145/3626772.3657707} to directly predict satisfaction scores.
\textbf{Chain-of-Thought}~\cite{3600270.3602070} prompts LLMs to reason before assigning a satisfaction score.
\textbf{Rubric-Based Evaluation}, adapted from Gunjal et al.~\cite{gunjal2025rubricsrewardsreinforcementlearning}, first generates a rubric solely from the query without access to the AI overview and then uses an LLM judge to predict user satisfaction by assessing the AI overview against the generated criteria.
Finally, we reimplement G-Eval~\cite{liu-etal-2023-g}, which combines chain-of-thought prompting with a form-filling paradigm, for satisfaction prediction following Shi et al.~\cite{10.1145/3805712.3809934} and refer to this implementation as \textbf{G-Eval-Sat}. In our implementation, all these LLM-based baselines use DeepSeek-V3.2~\cite{deepseekai2025deepseekv32pushingfrontieropen}.

(3) \textbf{User-Centric Satisfaction Prediction Methods Incorporating User Behaviors.} First, in \textit{traditional web search}, user behaviors are incorporated into heuristic approaches and machine learning models to predict user satisfaction, including 
\textbf{Click-Reformulation Heuristic}, \textbf{Dwell-Time Heuristic}, \textbf{LSTM} and \textbf{Transformer} models, \textbf{MLP}, \textbf{Random Forest}, \textbf{Ridge}, and \textbf{GBDT}. To ensure a fair comparison, we include the AI overview quality score as an auxiliary input feature to these baselines.
 We train the machine learning baselines to regress user-reported satisfaction scores. Due to space limitations, we provide the input features used by these baselines in Appendix~\ref{sec:baseline}.
Second, in \textit{generative search}, \textbf{G-CoS}~\cite{10.1145/3805712.3809934} represents a pioneering effort to incorporate user behaviors into satisfaction prediction for ChatGPT-like systems. However, it treats user behavior solely as a cost and generated response quality as a gain. 
\vspace{-18pt}
\begin{table}[H]
\centering
\caption{Overall performance on the SearchLog26 dataset.}
\vspace{-10pt}
\resizebox{\linewidth}{!}{
\begin{tabular}{lllllll}
\hline
& Precision & Recall & F1-Score & P-$r$ & S-$\rho$ & MAE\\
\hline
\multicolumn{7}{l}{\textbf{AI Overview Quality Evaluation Methods}} \\
ARES (context relevance)~\cite{saad-falcon-etal-2024-ares} & 0.0376$^{**}$ & 0.1049$^{**}$ & 0.0553$^{**}$ & 0.1014$^{**}$ & 0.1013$^{**}$ & 2.3035$^{**}$\\
ARES (answer faithfulness)~\cite{saad-falcon-etal-2024-ares} & 0.0317$^{**}$ & 0.0182$^{**}$ & 0.0232$^{**}$ & 0.0807$^{**}$ & 0.0805$^{**}$ & 2.8566$^{**}$\\
ARES (answer relevance)~\cite{saad-falcon-etal-2024-ares} & 0.0295$^{**}$ & 0.0760$^{**}$ & 0.0425$^{**}$ & 0.2993$^{**}$ & 0.2979$^{**}$ & 2.3566$^{**}$\\
TruLens (context relevance)~\cite{ferrara2024ragtriad} & 0.0391$^{**}$ & 0.1251$^{**}$ & 0.0596$^{**}$ & 0.0911$^{**}$ & 0.1038$^{**}$ & 1.7039$^{**}$\\
TruLens (groundedness)~\cite{ferrara2024ragtriad}& 0.0499$^{**}$ & 0.2000$^{**}$ & 0.0799$^{**}$ & 0.0310$^{**}$ & 0.0469$^{**}$ & 1.8678$^{**}$\\
TruLens (answer relevance)~\cite{ferrara2024ragtriad} & 0.0498$^{**}$ & 0.2000$^{**}$ & 0.0798$^{**}$ & 0.3193$^{**}$ & 0.3160$^{**}$ & 1.4901$^{**}$\\
RAGAs (faithfulness) ~\cite{es-etal-2024-RAGAs}& 0.0442$^{**}$ & 0.0428$^{**}$ & 0.0435$^{**}$ & 0.0360$^{**}$ & 0.0205$^{**}$ & 2.7799$^{**}$\\
RAGAs (answer relevance)~\cite{es-etal-2024-RAGAs} & 0.0331$^{**}$ & 0.0887$^{**}$ & 0.0483$^{**}$ & 0.2324$^{**}$ & 0.2240$^{**}$ & 2.3773$^{**}$\\
RAGAs (context relevance)~\cite{es-etal-2024-RAGAs} & 0.0403$^{**}$ & 0.1325$^{**}$ & 0.0619$^{**}$ & 0.0094$^{**}$ & 0.0210$^{**}$ & 2.2212$^{**}$\\
OpenEval (faithfulness)~\cite{ispas2025towards}  & 0.0384$^{**}$ & 0.0572$^{**}$ & 0.0459$^{**}$ & 0.0441$^{**}$ & 0.0435$^{**}$ & 2.6275$^{**}$\\
\hline
\multicolumn{7}{l}{\textbf{AI Overview User-Centric Evaluation Methods}} \\
GAH (coherence)~\cite{10.1145/3626772.3657849} & 0.2275$^{**}$ & 0.2270$^{**}$ & 0.1169$^{**}$ & 0.2578$^{**}$ & 0.2389$^{**}$ & 1.3968$^{**}$\\
GAH (coverage)~\cite{10.1145/3626772.3657849} & \underline{0.3305} & 0.2461$^{**}$ & 0.1981$^{**}$ & 0.3163$^{**}$ & 0.2808$^{**}$ & 1.3244$^{**}$\\
GAH (consistency)~\cite{10.1145/3626772.3657849} & 0.2703$^{**}$ & 0.2210$^{**}$ & 0.1273$^{**}$ & 0.2042$^{**}$ & 0.1731$^{**}$ & 1.6028$^{**}$\\
GAH (correctness)~\cite{10.1145/3626772.3657849} & 0.2538$^{**}$ & 0.2356$^{**}$ & 0.1665$^{**}$ & 0.2851$^{**}$ & 0.2478$^{**}$ & 1.3407$^{**}$\\
GAH (clarity)~\cite{10.1145/3626772.3657849} & 0.2605$^{**}$ & 0.2204$^{**}$ & 0.1198$^{**}$ & 0.2599$^{**}$ & 0.2216$^{**}$ & 1.4325$^{**}$\\
GAH (overall)~\cite{10.1145/3626772.3657849} & 0.2197$^{**}$ & 0.2448$^{**}$ & 0.1719$^{**}$ & 0.3109$^{**}$ & 0.2768$^{**}$ & 1.2845$^{**}$\\
URS~\cite{wang-etal-2024-user} & 0.2795$^{**}$ & 0.2510$^{**}$ & 0.2131$^{**}$ & 0.3439$^{**}$ & 0.3315$^{**}$ & 1.3604$^{**}$\\
Direct Prompting~\cite{10.1145/3626772.3657707} & 0.2883$^{**}$ & 0.2547$^{**}$ & 0.2191$^{**}$ & 0.3463$^{**}$ & 0.3316$^{**}$ & 1.3558$^{**}$\\
Chain-of-Thought~\cite{3600270.3602070} & 0.2840$^{**}$ & 0.2493$^{**}$ & 0.2150$^{**}$  & 0.3355$^{**}$ & 0.3195$^{**}$ & 1.3625$^{**}$\\
Rubric-Based Evaluation~\cite{gunjal2025rubricsrewardsreinforcementlearning} & 0.2770$^{**}$ & \underline{0.2593} & 0.2308$^{**}$ & 0.3367$^{**}$ & 0.3148$^{**}$ & 1.4727$^{**}$\\
G-Eval-Sat~\cite{liu-etal-2023-g} & 0.2565$^{**}$ & 0.2566$^{**}$ & 0.2234$^{**}$ & 0.3691$^{**}$ & 0.3621$^{**}$ & 1.4340$^{**}$\\
\hline
\multicolumn{7}{l}{\textbf{User-Centric Satisfaction Prediction Methods Incorporating User Behaviors}} \\
G-CoS~\cite{10.1145/3805712.3809934} & 0.2533$^{**}$ & 0.2316$^{**}$ & 0.2317$^{**}$ & 0.3328$^{**}$ & 0.3300$^{**}$ & 1.3126$^{**}$ \\
Click-Reformulation Heuristic & 0.2392$^{**}$ & 0.2343$^{**}$ & 0.2302$^{**}$ & 0.1774$^{**}$ & 0.1768$^{**}$ & 1.4480$^{**}$ \\
Dwell-Time Heuristic & 0.2595$^{**}$ & 0.2440$^{**}$ & \underline{0.2426}$^{**}$ & 0.2904$^{**}$ & 0.2819$^{**}$ & 1.3257$^{**}$ \\
Ridge & 0.2610$^{**}$ & 0.2587$^{**}$ & 0.2064$^{**}$ & 0.3993$^{**}$ & 0.3917$^{**}$ & 1.3010$^{**}$ \\
LSTM & 0.3111$^{**}$ & 0.2410$^{**}$ & 0.2194$^{**}$ & 0.3774$^{**}$ & 0.3841$^{**}$ & 1.3067$^{**}$ \\
Random Forest & 0.3025$^{**}$ & 0.2585$^{*}$ & 0.2251$^{**}$ & \underline{0.4072 }$^{**}$& 0.4054$^{**}$ & 1.2617$^{**}$ \\
MLP & 0.2802$^{**}$ & 0.2463$^{**}$ & 0.2361$^{**}$ & 0.3942$^{**}$ & 0.3817$^{**}$ & 1.2775$^{**}$ \\
Transformer & 0.1995$^{**}$ & 0.2402$^{**}$ & 0.1859$^{**}$ & 0.3747$^{**}$ & 0.3776$^{**}$ & 1.2527$^{**}$ \\
GBDT & 0.3230$^{**}$ & 0.2565$^{**}$ & 0.2041$^{**}$ & 0.4053$^{**}$ & \underline{0.4081 }$^{**}$& \underline{1.2291}$^{**}$ \\
\hline
STaRS & \textbf{0.3832} & \textbf{0.2931} & \textbf{0.2580} & \textbf{0.4638} & \textbf{0.4652} & \textbf{1.0844}\\
\hline
\end{tabular}
}
\par\vspace{2pt}
\begin{minipage}{\linewidth}
\footnotesize
$^{*}p < 0.05$, $^{**}p < 0.01$.
Stars on baseline entries indicate that STaRS is significantly better on that metric (sample-level paired bootstrap test, \(B = 10{,}000\) resamples).
\end{minipage}
\vspace{-14pt}
\label{tab:overall_qiyeshuju}
\end{table}
\vspace{-14pt}
\subsection{Evaluation Metrics}
We employ two types of metrics to evaluate predicted scores and user-reported scores.

\textbf{Classification Metrics:}
Treating satisfaction scores as classes, we compute macro-averaged Precision (P), Recall (R), and F1-score (F1).
\textbf{Consistency Metrics:}
Treating satisfaction scores as numerical values, we assess correlations between predicted and user-reported scores using Pearson's $r$ (P-$r$) and Spearman's $\rho$ (S-$\rho$), and quantify prediction error using mean absolute error (MAE).
\vspace{-8pt}
\subsection{Overall Performance}
We evaluate our method on two datasets with different scales and experimental settings, so we will analyze them separately.

\paragraph{UserStudy Dataset.}
Given the limited size of the UserStudy dataset, we are unable to apply RL and must therefore rely solely on the structured reasoning strategy component of STaRS. 
We use DeepSeek-V3.2 to execute the structured reasoning strategy component of STaRS without task-specific fine-tuning. All LLM-based baselines in this experiment also use DeepSeek-V3.2.
All metrics reported in the table are computed over the entire UserStudy dataset. For STaRS and baselines that do not require task-specific training, we directly use the entire UserStudy dataset as the test set. For the baselines in the third category that require task-specific training, we split the data into training and test sets by query and use five-fold cross-validation. The two heuristics require no training and use thresholds set from prior experience.
The results are shown in Table~\ref{tab:overall_userstudy}.
We observe the following: (1)
Within each AI overview quality evaluation method, answer relevance shows the strongest correlation with user-reported satisfaction, which suggests that it have a greater influence on user satisfaction than other metrics, such as context relevance. (2) STaRS achieves better overall performance than all AI overview user-centric evaluation methods, which highlight the limitations of predicting satisfaction from the query and AI overview alone.
(3) STaRS outperforms strong machine learning baselines in the third category, which suggests that sota satisfaction prediction methods in traditional IR have limited effectiveness when  applied to AI overview satisfaction prediction and demonstrates the effectiveness of STaRS. (4) STaRS without RL fine-tuning achieves the best overall performance across the three categories, which supports the effectiveness of the structured reasoning strategy without task-specific training. This also shows that in settings with limited data, i.e., when companies have not collected enough real user satisfaction feedback for model training, our method can achieve good performance and offer practical value.
\vspace{-8pt}
\paragraph{SearchLog26 Dataset.} On the larger commercial SearchLog26 dataset, we evaluate the full STaRS method, which incorporates both the structured reasoning strategy and RL with a reward shaping mechanism. We use Qwen3-8B~\cite{yang2025qwen3technicalreport}  as the backbone of STaRS.
The RL setup of STaRS is described in Section~\ref{sec:details_RL}.
All LLM-based baselines in this experiment use DeepSeek-V3.2. Following prior RL-based IR work~\cite{zeng2025optimizing,chen2025toprankedrecommendationsfailmodeling}, we adopt a query-level holdout protocol by randomly splitting the data into 5,271 training queries (80.00\%) and 1,318 test queries (20.00\%). All queries from the same user are assigned to the same split. All methods are evaluated on the same test set. 
From the results in Table~\ref{tab:overall_qiyeshuju}, we can observe:
(1) Scores produced by AI overview quality evaluation methods generally show weak correlation with user-reported satisfaction on SearchLog26, which suggests that system-centric evaluation methods struggle to achieve the user-centric evaluation objective.
(2) STaRS outperforms all AI overview user-centric evaluation methods, which shows that AI overview content alone is insufficient to capture satisfaction.
(3) STaRS outperforms two heuristic methods, which suggests that existing and fixed assumptions about the satisfaction implications of user behavior cannot transfer directly to AI overview satisfaction prediction.
(4) Satisfaction prediction methods incorporating user behaviors generally perform best among the three categories, with STaRS leading this group, which supports both the value of behavior signals and the effectiveness of our method.
(5) STaRS outperforms G-CoS~\cite{10.1145/3805712.3809934}, which suggests that G-CoS's treatment of user behavior solely as a cost, originally proposed for ChatGPT-like systems, may not fully capture the implications of behavior signals for AI overview satisfaction prediction. (6) STaRS with RL fine-tuning achieves the best overall performance across the three categories. This shows that with sufficient user satisfaction feedback, our method outperforms other models trained on the same amount of data, demonstrating its practical value in enterprise.

\vspace{-8pt}
\subsection{Ablation Studies}
We conduct ablation studies to assess the impact of key STaRS designs from the following two perspectives:

\textbf{Effectiveness of the Structured Reasoning Strategy and Inputs.} To demonstrate the importance of different input information for predicting satisfaction, 
we remove $a$, $B$, or $C$ from the input tuple $(q,a,B,C)$ one at a time. To examine the effect of the structured reasoning strategy, we replace our instruction with one that directly predicts the satisfaction score. Given the cost of RL training, we use DeepSeek-V3.2 in a zero-shot setting without RL training for all variants on both datasets. On UserStudy and SearchLog26, all variants are evaluated on the full dataset and the same 1,318-query test set used in the main experiment, respectively. 

As shown in Table~\ref{tab:ablation1}, removing behavior signals reduces both Pearson and Spearman correlations and increases MAE on both datasets. On both datasets, the resulting performance is close to that of user-centric AI overview evaluation methods, which do not incorporate behavior signals (Tables~\ref{tab:overall_userstudy} and~\ref{tab:overall_qiyeshuju}). These results support the value of observed user interactions as input for satisfaction prediction beyond response content.
Removing AI overview content or context information yields the same pattern, supporting the contribution of each input within the full STaRS model.
Furthermore, replacing the structured reasoning strategy with direct prediction also lowers correlations and increases MAE, supporting the value of a fine-grained, behavior-signal-wise reasoning strategy, despite its lower macro-F1 on UserStudy.
\vspace{-2pt}
\begin{table}[H]
\centering
\vspace{-8pt}
\caption{Ablation study of inputs and the structured reasoning strategy for AI overview satisfaction prediction.}
\vspace{-10pt}
\label{tab:ablation1}
\resizebox{\linewidth}{!}{
\begin{tabular}{lcccccc}
\hline
 & Precision & Recall & F1-Score & P-$r$ & S-$\rho$ & MAE \\
\hline
\textbf{SearchLog26 Dataset} & & & & & & \\
All Inputs + Structured Reasoning Strategy
& 0.3048 & \textbf{0.2877} & \textbf{0.2667} & \textbf{0.4120} & \textbf{0.4029} & \textbf{1.1593 }\\
w/o Content $a$
& \textbf{0.4362} & 0.2425 & 0.1571 & 0.3615 & 0.3530 & 1.2148 \\

w/o Behavior Signals $B$
& 0.2796 & 0.2465 & 0.2048 & 0.3500 & 0.3327 & 1.4356 \\

w/o Context Information $C$
& 0.3315 & 0.2448 & 0.1991 & 0.3813 & 0.3788 & 1.3303 \\

w/o Structured Reasoning Strategy
& 0.3197 & 0.2400 & 0.1915 & 0.3582 & 0.3551 & 1.3336 \\

\hline
\textbf{UserStudy Dataset} & & & & & & \\
All Inputs + Structured Reasoning Strategy
& 0.3833 & 0.2783 & 0.2650 & \textbf{0.4051} & \textbf{0.4157 }& \textbf{0.8646} \\

w/o Content $a$
& 0.1323 & 0.2033 & 0.1386 & 0.1650 & 0.1551 & 1.1733 \\

w/o Behavior Signals $B$
& 0.2555 & 0.2317 & 0.2215 & 0.2866 & 0.3054 & 0.9657 \\

w/o Context Information $C$
& \textbf{0.4562} & 0.2726 & 0.2689 & 0.3809 & 0.3881 & 0.8881 \\
w/o Structured Reasoning Strategy
& 0.3530 & \textbf{0.2875} & \textbf{0.2892} & 0.3804 & 0.3807 & 0.9134 \\

\hline
\end{tabular}
}
\vspace{-4pt}
\end{table}
\textbf{Effectiveness of RL and Reward Shaping.} We examine the contributions of RL optimization and our proposed reward shaping mechanism through the following four variants:
(1) \textit{w/o Counterfactual Signal Selection} removes the counterfactual signal selection stage in Section~\ref{sec:reward-shaping}, that is, $R_p$ is computed using $P_{\mathrm{ref}}^{\mathrm{vote}}$ instead of $P_{\mathrm{ref}}$;
(2) \textit{w/o Reward Shaping} removes the behavior-signal-level supervision in Section~\ref{sec:reward-shaping} and the corresponding reward $R_p$ in Section~\ref{sec:rl-rewards};
(3) \textit{w/o RL} removes the RL optimization in Section~\ref{sec:rl-optimization} and directly uses the original LLM checkpoint without task-specific fine-tuning;
and (4) \textit{SFT instead of RL} replaces the RL optimization in Section~\ref{sec:rl-optimization} with supervised fine-tuning. 

We conduct these ablations on SearchLog26. We use the same SearchLog26 train/test split as in the main experiment. The backbones used in this experiment are Qwen3-8B~\cite{yang2025qwen3technicalreport} and Llama-3-8B-Instruct~\cite{grattafiori2024llama3herdmodels}. 
For the RL-trained variants, we use the same RL setup as in the main experiments (Section~\ref{sec:details_RL}).
For the SFT-trained variants, we perform full-parameter fine-tuning with response-only cross-entropy on ground-truth satisfaction scores formatted as \texttt{<answer>score</answer>}, using AdamW for 2 epochs.

As shown in Table~\ref{tab:ablation}, all ablated variants exhibit lower overall performance than the full STaRS model on both Qwen and Llama.
Specifically, \textit{w/o Counterfactual Signal Selection} rewards peripheral and redundant judgments alongside influential ones, diluting supervision of behavior signals that are strongly associated with satisfaction prediction.
In \textit{w/o Reward Shaping}, the model relies on sparse outcome rewards, which evaluate final predictions without assessing reasoning quality and thus cannot distinguish sound reasoning traces from flawed ones that happen to produce correct predictions.
Moreover, \textit{w/o RL} highlights the role of RL optimization in adapting compact pretrained models to satisfaction prediction. Compared with SFT, RL achieves better performance, suggesting stronger reasoning and generalization rather than relying solely on input–label associations learned during supervised training.
\vspace{-10pt}
\begin{table}[H]
\centering
\caption{Ablation study of RL and reward shaping in STaRS on SearchLog26 using Qwen and Llama as backbone models.}
\vspace{-10pt}
\label{tab:ablation}
\resizebox{\linewidth}{!}{
\begin{tabular}{lcccccc}
\hline
Variant & Precision & Recall & F1-Score & P-$r$ & S-$\rho$ & MAE \\
\hline
\multicolumn{7}{l}{\textbf{Qwen}} \\
\hline
STaRS (Full Model)            & \textbf{0.3832} & \textbf{0.2931} & \textbf{0.2580} & \textbf{0.4638} & \textbf{0.4652} & \textbf{1.0844} \\
w/o Counterfactual Signal Selection   & 0.3433 & 0.2767 & 0.2441 & 0.4268 & 0.4360 & 1.1287 \\
w/o Reward Shaping            & 0.3485 & 0.2683 & 0.2307 & 0.4113 & 0.4026 & 1.1418 \\
w/o RL                        & 0.2900 & 0.2270 & 0.1712 & 0.2691 & 0.3033 & 1.2008 \\
SFT instead of RL             & 0.2992 & 0.2405 & 0.2334 & 0.3496 & 0.3502 & 1.1564 \\
\hline
\multicolumn{7}{l}{\textbf{Llama}} \\
\hline
STaRS (Full Model)            & 0.3524 & \textbf{0.2653} & \textbf{0.2303} & \textbf{0.3945} & \textbf{0.3866} & \textbf{1.1649} \\
w/o Counterfactual Signal Selection   & \textbf{0.3540} & 0.2621 & 0.2242 & 0.3781 & 0.3568 & 1.1789 \\
w/o Reward Shaping            & 0.3395 & 0.2579 & 0.2192 & 0.3419 & 0.3220 & 1.1999 \\
w/o RL                        & 0.3279 & 0.2280 & 0.1719 & 0.2366 & 0.2829 & 1.3680 \\
SFT instead of RL             & 0.3238 & 0.2395 & 0.2137 & 0.3414 & 0.3348 & 1.2299 \\
\hline
\end{tabular}
}
\vspace{-14pt}
\end{table}

\vspace{-4pt}
\section{Conclusion and Future Work}
System-centric evaluation cannot fully capture user satisfaction with AI overviews, while existing evaluation methods largely overlook informative user behavior.
In this work, we study the task of predicting  satisfaction with AI overviews by incorporating user behavior signals; to this end, we collect two new datasets to support this research: a high-quality laboratory dataset and a dataset from real-world search platforms. Based on the collected user-reported satisfaction, we derive several insights: quality evaluation methods show relatively weak correlations with user satisfaction with AI overviews, while incorporating user behavior further improves satisfaction prediction compared with using only the AI overview content as input.
We propose STaRS to predict AI overview satisfaction by leveraging the reasoning capabilities of LLMs. Designed specifically for AI overview satisfaction prediction, STaRS introduces two innovative components: a structured reasoning strategy and a reward shaping mechanism. Experiments show that STaRS outperforms AI overview evaluation methods and existing behavior-based satisfaction prediction methods in traditional IR. These strengths offer search providers a practical approach to accurately predicting user satisfaction with AI overviews. 

Two promising directions for future work are (1) to use predicted satisfaction as a direct reward signal for optimizing AI overview systems using RL and (2) to leverage identified causes of dissatisfaction for more targeted improvements to AI overview systems. We hope our work can make a contribution to improving the user experience of AI overviews and that our new datasets, insights, and method contribute to further research in this area.
\vspace{-8pt}
\section{Ethical Use of Data}
We used anonymized search interaction records and satisfaction scores from a controlled user study and an industrial search platform. Participants in the user study provided informed consent for the collection, research use, and publication of their anonymized data. The industrial data were collected and used in accordance with the platform's privacy and data-use policies. This study involved only routine search activities and satisfaction ratings and did not cause any physical or psychological harm to users.

\bibliographystyle{ACM-Reference-Format}
\bibliography{sample-base}

@inproceedings{zeng2025optimizing,
  title={Optimizing Generative Ranking Relevance via Reinforcement Learning in Xiaohongshu Search},
  author={Zeng, Ziyang and Jing, Heming and Chen, Jindong and Li, Xiangli and Liu, Hongyu and He, Yixuan and Li, Zhengyu and Sun, Yige and Xie, Zheyong and Yang, Yuqing and others},
  booktitle={Proceedings of the 32nd ACM SIGKDD Conference on Knowledge Discovery and Data Mining V. 1},
  pages={2551--2561},
  year={2026}
}

@inproceedings{kiseleva2016predicting,
  title={Predicting user satisfaction with intelligent assistants},
  author={Kiseleva, Julia and Williams, Kyle and Hassan Awadallah, Ahmed and Crook, Aidan C and Zitouni, Imed and Anastasakos, Tasos},
  booktitle={Proceedings of the 39th International ACM SIGIR conference on Research and Development in Information Retrieval},
  pages={45--54},
  year={2016}
}

@inproceedings{10.1145/3209978.3210059,
author = {Zhang, Fan and Zhou, Ke and Shao, Yunqiu and Luo, Cheng and Zhang, Min and Ma, Shaoping},
title = {How Well do Offline and Online Evaluation Metrics Measure User Satisfaction in Web Image Search?},
year = {2018},
isbn = {9781450356572},
publisher = {Association for Computing Machinery},
address = {New York, NY, USA},
url = {https://doi.org/10.1145/3209978.3210059},
doi = {10.1145/3209978.3210059},
booktitle = {The 41st International ACM SIGIR Conference on Research \& Development in Information Retrieval},
pages = {615–624},
numpages = {10},
location = {Ann Arbor, MI, USA},
series = {SIGIR '18}
}

@inproceedings{10.1145/3077136.3080804,
author = {Chen, Ye and Zhou, Ke and Liu, Yiqun and Zhang, Min and Ma, Shaoping},
title = {Meta-evaluation of Online and Offline Web Search Evaluation Metrics},
year = {2017},
isbn = {9781450350228},
publisher = {Association for Computing Machinery},
address = {New York, NY, USA},
url = {https://doi.org/10.1145/3077136.3080804},
doi = {10.1145/3077136.3080804},
booktitle = {Proceedings of the 40th International ACM SIGIR Conference on Research and Development in Information Retrieval},
pages = {15–24},
numpages = {10},
location = {Shinjuku, Tokyo, Japan},
series = {SIGIR '17}
}

@inproceedings{liu-etal-2023-g,
    title = "{G}-Eval: {NLG} Evaluation using Gpt-4 with Better Human Alignment",
    author = "Liu, Yang  and
      Iter, Dan  and
      Xu, Yichong  and
      Wang, Shuohang  and
      Xu, Ruochen  and
      Zhu, Chenguang",
    editor = "Bouamor, Houda  and
      Pino, Juan  and
      Bali, Kalika",
    booktitle = "Proceedings of the 2023 Conference on Empirical Methods in Natural Language Processing",
    month = dec,
    year = "2023",
    address = "Singapore",
    publisher = "Association for Computational Linguistics",
    url = "https://aclanthology.org/2023.emnlp-main.153/",
    doi = "10.18653/v1/2023.emnlp-main.153",
    pages = "2511--2522"
}

@inproceedings{wang-etal-2024-user,
    title = "A User-Centric Multi-Intent Benchmark for Evaluating Large Language Models",
    author = "Wang, Jiayin  and
      Mo, Fengran  and
      Ma, Weizhi  and
      Sun, Peijie  and
      Zhang, Min  and
      Nie, Jian-Yun",
    editor = "Al-Onaizan, Yaser  and
      Bansal, Mohit  and
      Chen, Yun-Nung",
    booktitle = "Proceedings of the 2024 Conference on Empirical Methods in Natural Language Processing",
    month = nov,
    year = "2024",
    address = "Miami, Florida, USA",
    publisher = "Association for Computational Linguistics",
    url = "https://aclanthology.org/2024.emnlp-main.210/",
    doi = "10.18653/v1/2024.emnlp-main.210",
    pages = "3588--3612"
}

@inproceedings{saad-falcon-etal-2024-ares,
    title = "{ARES}: An Automated Evaluation Framework for Retrieval-Augmented Generation Systems",
    author = "Saad-Falcon, Jon  and
      Khattab, Omar  and
      Potts, Christopher  and
      Zaharia, Matei",
    editor = "Duh, Kevin  and
      Gomez, Helena  and
      Bethard, Steven",
    booktitle = "Proceedings of the 2024 Conference of the North American Chapter of the Association for Computational Linguistics: Human Language Technologies (Volume 1: Long Papers)",
    month = jun,
    year = "2024",
    address = "Mexico City, Mexico",
    publisher = "Association for Computational Linguistics",
    url = "https://aclanthology.org/2024.naacl-long.20/",
    doi = "10.18653/v1/2024.naacl-long.20",
    pages = "338--354"
}

@inproceedings{10.1145/3805712.3809934,
author = {Shi, Jia-Ling and Wu, Zhijing and Liang, Yidong and Mao, Xian-Ling},
title = {G-CoS: An Interpretable Gain-Cost Framework for User Satisfaction Estimation in Generative Information Retrieval},
year = {2026},
isbn = {9798400725999},
publisher = {Association for Computing Machinery},
address = {New York, NY, USA},
url = {https://doi.org/10.1145/3805712.3809934},
doi = {10.1145/3805712.3809934},
booktitle = {Proceedings of the 49th International ACM SIGIR Conference on Research and Development in Information Retrieval},
pages = {4110–4115},
numpages = {6},
location = {Australia},
series = {SIGIR '26}
}

@inproceedings{10.1145/3626772.3657849,
author = {Gienapp, Lukas and Scells, Harrisen and Deckers, Niklas and Bevendorff, Janek and Wang, Shuai and Kiesel, Johannes and Syed, Shahbaz and Fr{\"o}be, Maik and Zuccon, Guido and Stein, Benno and Hagen, Matthias and Potthast, Martin},
title = {Evaluating Generative Ad Hoc Information Retrieval},
year = {2024},
isbn = {9798400704314},
publisher = {Association for Computing Machinery},
address = {New York, NY, USA},
url = {https://doi.org/10.1145/3626772.3657849},
doi = {10.1145/3626772.3657849},
booktitle = {Proceedings of the 47th International ACM SIGIR Conference on Research and Development in Information Retrieval},
pages = {1916–1929},
numpages = {14},
location = {Washington DC, USA},
series = {SIGIR '24}
}

@inproceedings{es-etal-2024-ragas,
    title = "{RAGA}s: Automated Evaluation of Retrieval Augmented Generation",
    author = "Es, Shahul  and
      James, Jithin  and
      Espinosa Anke, Luis  and
      Schockaert, Steven",
    editor = "Aletras, Nikolaos  and
      De Clercq, Orphee",
    booktitle = "Proceedings of the 18th Conference of the European Chapter of the Association for Computational Linguistics: System Demonstrations",
    month = mar,
    year = "2024",
    address = "St. Julians, Malta",
    publisher = "Association for Computational Linguistics",
    url = "https://aclanthology.org/2024.eacl-demo.16/",
    doi = "10.18653/v1/2024.eacl-demo.16",
    pages = "150--158"
}

@inproceedings{ispas2025towards,
  title={Towards Lighter and Robust Evaluation for Retrieval Augmented Generation},
  author={Alex-R{\u{a}}zvan Ispas and Charles-{\'E}lie Simon and Fabien Caspani and Vincent Guigue},
  booktitle={ICLR Workshop: Quantify Uncertainty and Hallucination in Foundation Models: The Next Frontier in Reliable AI},
  year={2025},
  url={https://openreview.net/forum?id=3XWwf65dFW},
  eprint={2503.16161},
  archivePrefix={arXiv},
  primaryClass={cs.CL}
}

@misc{ferrara2024ragtriad,
  author       = {{TruLens}},
  title        = {The {RAG} Triad},
  howpublished = {\url{https://www.trulens.org/getting_started/core_concepts/rag_triad/}},
  note         = {Accessed: 2026-10-03}
}

@inproceedings{white2012leaving,
author = {Diriye, Abdigani and White, Ryen and Buscher, Georg and Dumais, Susan},
title = {Leaving so soon? understanding and predicting web search abandonment rationales},
year = {2012},
isbn = {9781450311564},
publisher = {Association for Computing Machinery},
address = {New York, NY, USA},
url = {https://doi.org/10.1145/2396761.2398399},
doi = {10.1145/2396761.2398399},
booktitle = {Proceedings of the 21st ACM International Conference on Information and Knowledge Management},
pages = {1025–1034},
numpages = {10},
location = {Maui, Hawaii, USA},
series = {CIKM '12}
}

@inproceedings{10.1145/2983323.2983829,
author = {Chuklin, Aleksandr and de Rijke, Maarten},
title = {Incorporating Clicks, Attention and Satisfaction into a Search Engine Result Page Evaluation Model},
year = {2016},
isbn = {9781450340731},
publisher = {Association for Computing Machinery},
address = {New York, NY, USA},
url = {https://doi.org/10.1145/2983323.2983829},
doi = {10.1145/2983323.2983829},
booktitle = {Proceedings of the 25th ACM International on Conference on Information and Knowledge Management},
pages = {175–184},
numpages = {10},
location = {Indianapolis, Indiana, USA},
series = {CIKM '16}
}

@inproceedings{hassan2013beyond,
  title={Beyond clicks: query reformulation as a predictor of search satisfaction},
  author={Hassan, Ahmed and Shi, Xiaolin and Craswell, Nick and Ramsey, Bill},
  booktitle={Proceedings of the 22nd ACM international conference on Information \& Knowledge Management},
  pages={2019--2028},
  year={2013}
}

@inproceedings{mehrotra2017user,
  title={User interaction sequences for search satisfaction prediction},
  author={Mehrotra, Rishabh and Zitouni, Imed and Hassan Awadallah, Ahmed and Kholy, Ahmed El and Khabsa, Madian},
  booktitle={Proceedings of the 40th International ACM SIGIR conference on research and development in information retrieval},
  pages={165--174},
  year={2017}
}

@article{fox2005evaluating,
  title={Evaluating implicit measures to improve web search},
  author={Fox, Steve and Karnawat, Kuldeep and Mydland, Mark and Dumais, Susan and White, Thomas},
  journal={ACM Transactions on Information Systems (TOIS)},
  volume={23},
  number={2},
  pages={147--168},
  year={2005},
  publisher={ACM New York, NY, USA}
}

@inproceedings{chakraborty2014correlation,
  title={On correlation of absence time and search effectiveness},
  author={Chakraborty, Sunandan and Radlinski, Filip and Shokouhi, Milad and Baecke, Paul},
  booktitle={Proceedings of the 37th international ACM SIGIR conference on Research \& development in information retrieval},
  pages={1163--1166},
  year={2014}
}

@inproceedings{xu2009evaluating,
  title={Evaluating web search using task completion time},
  author={Xu, Ya and Mease, David},
  booktitle={Proceedings of the 32nd international ACM SIGIR conference on Research and development in information retrieval},
  pages={676--677},
  year={2009}
}

@article{chen2017user,
  title={User satisfaction prediction with mouse movement information in heterogeneous search environment},
  author={Chen, Ye and Liu, Yiqun and Zhang, Min and Ma, Shaoping},
  journal={IEEE Transactions on Knowledge and Data Engineering},
  volume={29},
  number={11},
  pages={2470--2483},
  year={2017},
  publisher={IEEE}
}

@inproceedings{huang2012improving,
  title={Improving searcher models using mouse cursor activity},
  author={Huang, Jeff and White, Ryen W and Buscher, Georg and Wang, Kuansan},
  booktitle={Proceedings of the 35th international ACM SIGIR conference on Research and development in information retrieval},
  pages={195--204},
  year={2012}
}

@misc{bailey2015information,
  author       = {Bailey, Peter and Moffat, Alistair and Scholer, Falk and Thomas, Paul},
  title        = {Information Needs for {TREC} 2002--4 (2014)},
  year         = {2015},
  publisher    = {CSIRO},
  howpublished = {Data Collection, Version 2},
  doi          = {10.4225/08/55D0B6A098248},
  url          = {https://doi.org/10.4225/08/55D0B6A098248}
}

@inproceedings{lagun2014towards,
  title={Towards better measurement of attention and satisfaction in mobile search},
  author={Lagun, Dmitry and Hsieh, Chih-Hung and Webster, Dale and Navalpakkam, Vidhya},
  booktitle={Proceedings of the 37th international ACM SIGIR conference on Research \& development in information retrieval},
  pages={113--122},
  year={2014}
}

@inproceedings{xia2026search,
  title={Search-p1: Path-centric reward shaping for stable and efficient agentic rag training},
  author={Xia, Tianle and Xu, Ming and Hu, Lingxiang and Sun, Yiding and Li, Wenwei and Shang, Linfang and Liu, Liqun and Shu, Peng and Yu, Huan and Jiang, Jie},
  booktitle={Proceedings of the 64th Annual Meeting of the Association for Computational Linguistics (ACL 2026)},
  pages={517--534},
  year={2026}
}

@article{choi2025mentor,
  title={MENTOR: Reinforcement Learning via Flexible Teacher-Optimized Rewards for Tool-Use Distillation},
  author={Choi, ChangSu and Song, Hoyun and Kim, Dongyeon and Cho, Minkyung and Jung, WooHyeon and Park, Sunjin and Bae, NohHyeob and Yu, Seona and Lim, KyungTae},
  journal={arXiv preprint arXiv:2510.18383},
  year={2025}
}

@inproceedings{zheng2026comprehensive,
  title={A Comprehensive Survey of Process Reward Models: Data Generation, Model Construction, and Usage},
  author={Zheng, Congmin and Zhu, Jiachen and Ou, Zhuoying and Chen, Yuxiang and Zhang, Kangning and Shan, Rong and Zheng, Zeyu and Yang, Mengyue and Lin, Jianghao and Yu, Yong and others},
  booktitle={Proceedings of the 64th Annual Meeting of the Association for Computational Linguistics (Volume 1: Long Papers)},
  pages={3591--3607},
  year={2026}
}

@article{yu2025reward,
  title={Reward models in deep reinforcement learning: A survey},
  author={Yu, Rui and Wan, Shenghua and Wang, Yucen and Gao, Chen-Xiao and Gan, Le and Zhang, Zongzhang and Zhan, De-Chuan},
  journal={arXiv preprint arXiv:2506.15421},
  year={2025}
}

@article{schoepp2025evolving,
  title={The evolving landscape of llm-and vlm-integrated reinforcement learning},
  author={Schoepp, Sheila and Jafaripour, Masoud and Cao, Yingyue and Yang, Tianpei and Abdollahi, Fatemeh and Golestan, Shadan and Sufiyan, Zahin and Zaiane, Osmar R and Taylor, Matthew E},
  journal={arXiv preprint arXiv:2502.15214},
  year={2025}
}

@misc{shao2024deepseekmathpushinglimitsmathematical,
      title={DeepSeekMath: Pushing the Limits of Mathematical Reasoning in Open Language Models}, 
      author={Zhihong Shao and Peiyi Wang and Qihao Zhu and Runxin Xu and Junxiao Song and Xiao Bi and Haowei Zhang and Mingchuan Zhang and Y. K. Li and Y. Wu and Daya Guo},
      year={2024},
      eprint={2402.03300},
      archivePrefix={arXiv},
      primaryClass={cs.CL},
      url={https://arxiv.org/abs/2402.03300}, 
}

@inproceedings{10.1145/1718487.1718515,
author = {Hassan, Ahmed and Jones, Rosie and Klinkner, Kristina Lisa},
title = {Beyond DCG: user behavior as a predictor of a successful search},
year = {2010},
isbn = {9781605588896},
publisher = {Association for Computing Machinery},
address = {New York, NY, USA},
url = {https://doi.org/10.1145/1718487.1718515},
doi = {10.1145/1718487.1718515},
booktitle = {Proceedings of the Third ACM International Conference on Web Search and Data Mining},
pages = {221–230},
numpages = {10},
location = {New York, New York, USA},
series = {WSDM '10}
}

@inproceedings{chen2025toprankedrecommendationsfailmodeling,
  title={When top-ranked recommendations fail: Modeling multi-granular negative feedback for explainable and robust video recommendation},
  author={Chen, Siran and Chen, Boyu and Yu, Chenyun and Ouyang, Yi and Cheng, Lei and Zhuo, Chengxiang and Li, Zang and Wang, Yali},
  booktitle={Proceedings of the AAAI Conference on Artificial Intelligence},
  volume={40},
  number={24},
  pages={20262--20270},
  year={2026}
}

@article{dan2016measuring,
  title={Measuring and predicting search engine users’ satisfaction},
  author={Dan, Ovidiu and Davison, Brian D},
  journal={ACM Computing Surveys (CSUR)},
  volume={49},
  number={1},
  pages={1--35},
  year={2016},
  publisher={ACM New York, NY, USA}
}

@inproceedings{jiang2015understanding,
  title={Understanding and predicting graded search satisfaction},
  author={Jiang, Jiepu and Hassan Awadallah, Ahmed and Shi, Xiaolin and White, Ryen W},
  booktitle={Proceedings of the eighth ACM international conference on web search and data mining},
  pages={57--66},
  year={2015}
}

@inproceedings{10.1145/2600428.2609607,
author = {Wang, Hongning and Song, Yang and Chang, Ming-Wei and He, Xiaodong and Hassan, Ahmed and White, Ryen W.},
title = {Modeling action-level satisfaction for search task satisfaction prediction},
year = {2014},
isbn = {9781450322577},
publisher = {Association for Computing Machinery},
address = {New York, NY, USA},
url = {https://doi.org/10.1145/2600428.2609607},
doi = {10.1145/2600428.2609607},
booktitle = {Proceedings of the 37th International ACM SIGIR Conference on Research \& Development in Information Retrieval},
pages = {123–132},
numpages = {10},
location = {Gold Coast, Queensland, Australia},
series = {SIGIR '14}
}

@inproceedings{10.1145/3336191.3371799,
author = {Wicaksono, Alfan Farizki and Moffat, Alistair},
title = {Metrics, User Models, and Satisfaction},
year = {2020},
isbn = {9781450368223},
publisher = {Association for Computing Machinery},
address = {New York, NY, USA},
url = {https://doi.org/10.1145/3336191.3371799},
doi = {10.1145/3336191.3371799},
booktitle = {Proceedings of the 13th International Conference on Web Search and Data Mining},
pages = {654–662},
numpages = {9},
location = {Houston, TX, USA},
series = {WSDM '20}
}

@inproceedings{zhang2020models,
author = {Zhang, Fan and Mao, Jiaxin and Liu, Yiqun and Xie, Xiaohui and Ma, Weizhi and Zhang, Min and Ma, Shaoping},
title = {Models Versus Satisfaction: Towards a Better Understanding of Evaluation Metrics},
year = {2020},
isbn = {9781450380164},
publisher = {Association for Computing Machinery},
address = {New York, NY, USA},
url = {https://doi.org/10.1145/3397271.3401162},
doi = {10.1145/3397271.3401162},
booktitle = {Proceedings of the 43rd International ACM SIGIR Conference on Research and Development in Information Retrieval},
pages = {379–388},
numpages = {10},
location = {Virtual Event, China},
series = {SIGIR '20}
}

@inproceedings{mao2016does,
  title={When does relevance mean usefulness and user satisfaction in web search?},
  author={Mao, Jiaxin and Liu, Yiqun and Zhou, Ke and Nie, Jian-Yun and Song, Jingtao and Zhang, Min and Ma, Shaoping and Sun, Jiashen and Luo, Hengliang},
  booktitle={Proceedings of the 39th International ACM SIGIR conference on Research and Development in Information Retrieval},
  pages={463--472},
  year={2016}
}

@inproceedings{10.1145/2766462.2767721,
author = {Liu, Yiqun and Chen, Ye and Tang, Jinhui and Sun, Jiashen and Zhang, Min and Ma, Shaoping and Zhu, Xuan},
title = {Different Users, Different Opinions: Predicting Search Satisfaction with Mouse Movement Information},
year = {2015},
isbn = {9781450336215},
publisher = {Association for Computing Machinery},
address = {New York, NY, USA},
url = {https://doi.org/10.1145/2766462.2767721},
doi = {10.1145/2766462.2767721},
booktitle = {Proceedings of the 38th International ACM SIGIR Conference on Research and Development in Information Retrieval},
pages = {493–502},
numpages = {10},
location = {Santiago, Chile},
series = {SIGIR '15}
}

@inproceedings{kim2014modeling,
  title={Modeling dwell time to predict click-level satisfaction},
  author={Kim, Youngho and Hassan, Ahmed and White, Ryen W and Zitouni, Imed},
  booktitle={Proceedings of the 7th ACM international conference on Web search and data mining},
  pages={193--202},
  year={2014}
}

@inproceedings{10.1145/2736277.2741669,
author = {Jiang, Jiepu and Hassan Awadallah, Ahmed and Jones, Rosie and Ozertem, Umut and Zitouni, Imed and Gurunath Kulkarni, Ranjitha and Khan, Omar Zia},
title = {Automatic Online Evaluation of Intelligent Assistants},
year = {2015},
isbn = {9781450334693},
publisher = {International World Wide Web Conferences Steering Committee},
address = {Republic and Canton of Geneva, CHE},
url = {https://doi.org/10.1145/2736277.2741669},
doi = {10.1145/2736277.2741669},
booktitle = {Proceedings of the 24th International Conference on World Wide Web},
pages = {506–516},
numpages = {11},
location = {Florence, Italy},
series = {WWW '15}
}

@inproceedings{10.1145/3292500.3330981,
author = {Liu, Mengyang and Mao, Jiaxin and Liu, Yiqun and Zhang, Min and Ma, Shaoping},
title = {Investigating Cognitive Effects in Session-level Search User Satisfaction},
year = {2019},
isbn = {9781450362016},
publisher = {Association for Computing Machinery},
address = {New York, NY, USA},
url = {https://doi.org/10.1145/3292500.3330981},
doi = {10.1145/3292500.3330981},
booktitle = {Proceedings of the 25th ACM SIGKDD International Conference on Knowledge Discovery \& Data Mining},
pages = {923–931},
numpages = {9},
location = {Anchorage, AK, USA},
series = {KDD '19}
}

@inproceedings{10.1145/3626772.3657707,
author = {Thomas, Paul and Spielman, Seth and Craswell, Nick and Mitra, Bhaskar},
title = {Large Language Models can Accurately Predict Searcher Preferences},
year = {2024},
isbn = {9798400704314},
publisher = {Association for Computing Machinery},
address = {New York, NY, USA},
url = {https://doi.org/10.1145/3626772.3657707},
doi = {10.1145/3626772.3657707},
booktitle = {Proceedings of the 47th International ACM SIGIR Conference on Research and Development in Information Retrieval},
pages = {1930–1940},
numpages = {11},
location = {Washington DC, USA},
series = {SIGIR '24}
}

@inproceedings{3600270.3602070,
author = {Wei, Jason and Wang, Xuezhi and Schuurmans, Dale and Bosma, Maarten and Ichter, Brian and Xia, Fei and Chi, Ed H. and Le, Quoc V. and Zhou, Denny},
title = {Chain-of-thought prompting elicits reasoning in large language models},
year = {2022},
isbn = {9781713871088},
publisher = {Curran Associates Inc.},
address = {Red Hook, NY, USA},
booktitle = {Proceedings of the 36th International Conference on Neural Information Processing Systems},
articleno = {1800},
numpages = {14},
location = {New Orleans, LA, USA},
series = {NIPS '22}
}

@inproceedings{gunjal2025rubricsrewardsreinforcementlearning,
  title={Rubrics as rewards: Reinforcement learning beyond verifiable domains},
  author={Gunjal, Anisha and Wang, Anthony and Lau, Elaine and Nath, Vaskar and He, Yunzhong and Liu, Bing and Hendryx, Sean},
  booktitle={International Conference on Learning Representations},
  volume={2026},
  pages={127924--127945},
  year={2026}
}

@inproceedings{3737916.3738608,
author = {Ru, Dongyu and Qiu, Lin and Hu, Xiangkun and Zhang, Tianhang and Shi, Peng and Chang, Shuaichen and Jiayang, Cheng and Wang, Cunxiang and Sun, Shichao and Li, Huanyu and Zhang, Zizhao and Wang, Binjie and Jiang, Jiarong and He, Tong and Wang, Zhiguo and Liu, Pengfei and Zhang, Yue and Zhang, Zheng},
title = {RAGCHECKER: a fine-grained framework for diagnosing retrieval-augmented generation},
year = {2024},
isbn = {9798331314385},
publisher = {Curran Associates Inc.},
address = {Red Hook, NY, USA},
booktitle = {Proceedings of the 38th International Conference on Neural Information Processing Systems},
articleno = {692},
numpages = {29},
location = {Vancouver, BC, Canada},
series = {NIPS '24}
}

@inproceedings{10.1145/3689031.3696075,
author = {Sheng, Guangming and Zhang, Chi and Ye, Zilingfeng and Wu, Xibin and Zhang, Wang and Zhang, Ru and Peng, Yanghua and Lin, Haibin and Wu, Chuan},
title = {HybridFlow: A Flexible and Efficient RLHF Framework},
year = {2025},
isbn = {9798400711961},
publisher = {Association for Computing Machinery},
address = {New York, NY, USA},
url = {https://doi.org/10.1145/3689031.3696075},
doi = {10.1145/3689031.3696075},
booktitle = {Proceedings of the Twentieth European Conference on Computer Systems},
pages = {1279–1297},
numpages = {19},
location = {Rotterdam, Netherlands},
series = {EuroSys '25}
}

@inproceedings{awadallah2013personalized,
  title={Personalized models of search satisfaction},
  author={Hassan, Ahmed and White, Ryen W},
  booktitle={Proceedings of the 22nd ACM international conference on Information \& Knowledge Management},
  pages={2009--2018},
  year={2013}
}

@misc{deepseekai2025deepseekv32pushingfrontieropen,
      title={DeepSeek-V3.2: Pushing the Frontier of Open Large Language Models}, 
      author={DeepSeek-AI and Aixin Liu and Aoxue Mei and others},
      year={2025},
      eprint={2512.02556},
      archivePrefix={arXiv},
      primaryClass={cs.CL},
      url={https://arxiv.org/abs/2512.02556}, 
}

@misc{yang2025qwen3technicalreport,
      title={Qwen3 Technical Report}, 
      author={An Yang and Anfeng Li and Baosong Yang and others},
      year={2025},
      eprint={2505.09388},
      archivePrefix={arXiv},
      primaryClass={cs.CL},
      url={https://arxiv.org/abs/2505.09388}, 
}

@misc{grattafiori2024llama3herdmodels,
      title={The Llama 3 Herd of Models}, 
      author={Aaron Grattafiori and Abhimanyu Dubey and Abhinav Jauhri and others},
      year={2024},
      eprint={2407.21783},
      archivePrefix={arXiv},
      primaryClass={cs.AI},
      url={https://arxiv.org/abs/2407.21783}, 
}

@article{singh2025openai,
  title={Openai gpt-5 system card},
  author={Singh, Aaditya and Fry, Adam and Perelman, Adam and Tart, Adam and Ganesh, Adi and El-Kishky, Ahmed and McLaughlin, Aidan and Low, Aiden and Ostrow, AJ and Ananthram, Akhila and others},
  journal={arXiv preprint arXiv:2601.03267},
  year={2025}
}

\appendix

\section{Effect of Reward Shaping Component on Intermediate Reasoning Quality}
\label{sec:reward_shaping_reasoning_quality}
The preceding experiments show that reward shaping improves satisfaction prediction accuracy. To examine whether reward shaping improves intermediate reasoning quality, we compare STaRS with its variant without reward shaping by evaluating their interpretations of user behavior signals against users’ subsequently reported reasons. Both models are trained on the same SearchLog26 training set using the same hyperparameters as in the main experiment (Appendix~\ref{sec:details_RL}). The \textit{w/o Reward Shaping} variant removes only the behavior-signal-level supervision and its corresponding reward $R_p$ during training. Both use Qwen3-8B as the backbone. Note that user-reported reasons are used only in this section, where they serve solely as ground-truth labels for evaluation; they are not used as model inputs or outputs in any experiments reported elsewhere in this paper.
\vspace{-6pt}
\subsection{Case Study}
\vspace{-10pt}
\begin{figure}[h]
    \centering
    \includegraphics[width=1\linewidth]{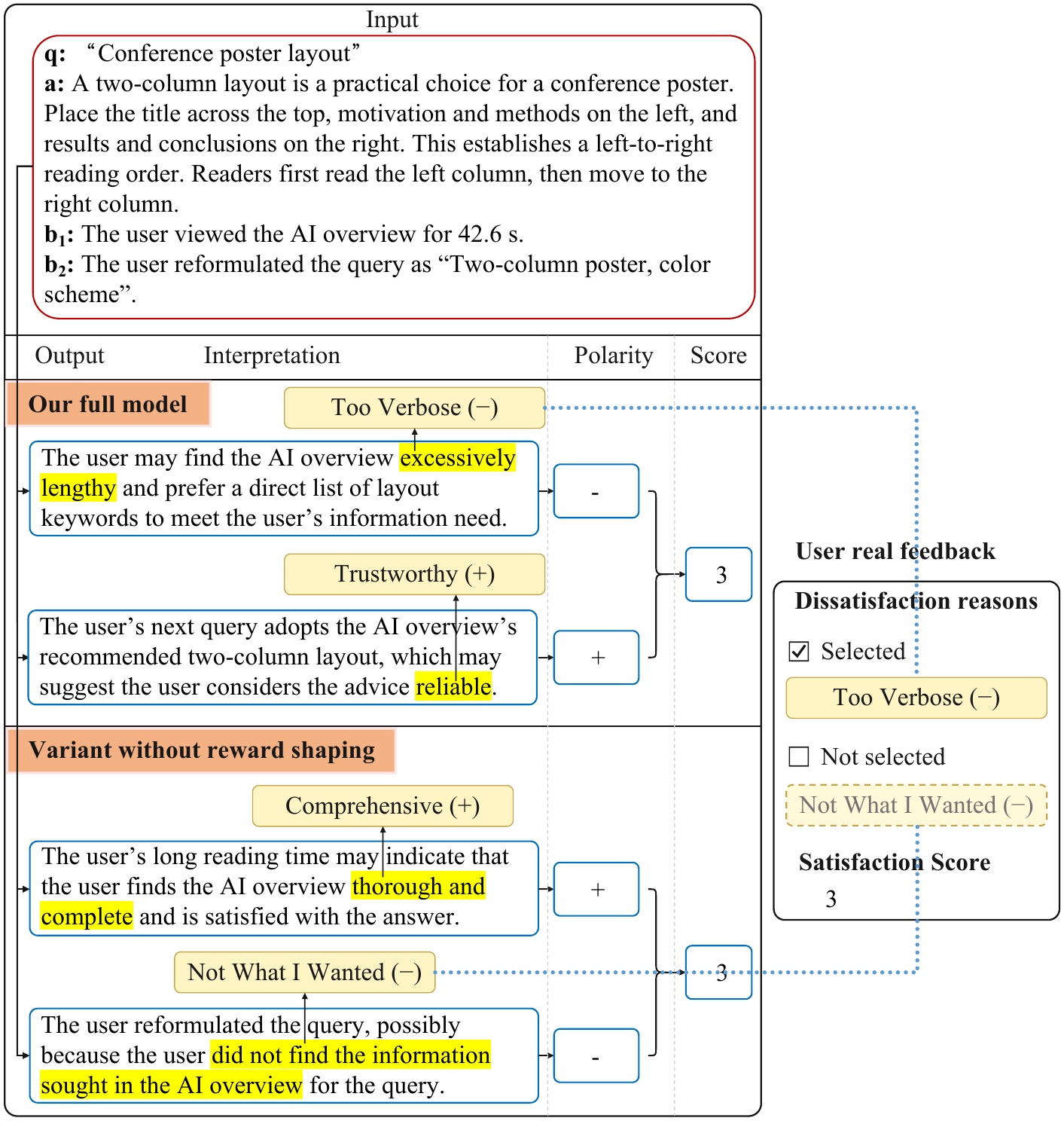}
    \vspace{-20pt}
    \caption{A case from the SearchLog26 dataset illustrating alignment between predicted interpretations $\hat  e$ and user-reported (dis)satisfaction reasons, comparing STaRS with its variant without reward shaping.}
    \label{fig:case}
    \vspace{-14pt}
\end{figure}

\begin{figure}[h]
    \centering
    \includegraphics[width=\linewidth]{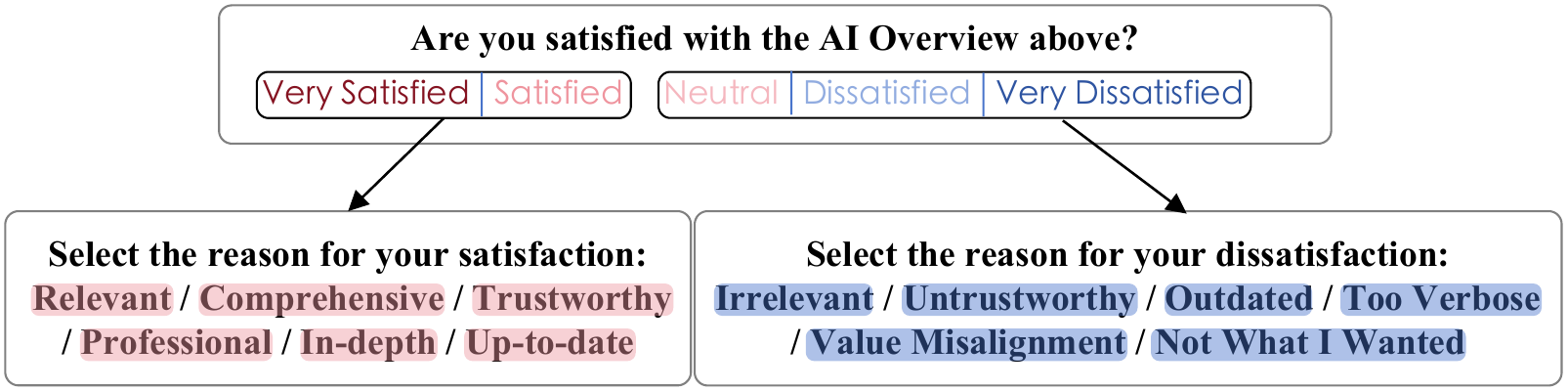}
    \vspace{-20pt}
    \caption{User feedback interface for collecting the AI overview satisfaction ratings and reasons recorded in the SearchLog26 dataset.}
    \vspace{-12pt}
\label{fig:app}
\end{figure}
Figure~\ref{fig:case} presents a case study, i.e., a qualitative comparison; red rectangles indicate the shared input, while blue rectangles show the output. Yellow labels summarize the reason categories expressed in the predicted interpretations. The right panel displays the user's actual feedback: checked boxes with solid yellow borders indicate selected reasons, whereas unchecked boxes with dashed yellow borders indicate unselected reasons.

In this example, the user searches for ``Conference poster layout'', views the AI overview for 42.6 seconds, and reformulates the query as ``Two-column poster, color scheme''. STaRS identifies the first behavior signal as potentially indicating that the user finds the AI overview excessively verbose and interprets it as a negative signal, while treating the adoption of the recommended layout as a positive signal. However, the variant without reward shaping associates viewing duration with comprehensiveness and reformulation with unmet information needs. As shown in Figure~\ref{fig:app}, users rate satisfaction on a five-point scale and select predefined reasons, with ratings of 4--5 triggering satisfaction reasons and 1--3 dissatisfaction reasons. The user selects ``Too Verbose'' but not ``Not What I Wanted''. Accordingly, STaRS captures the reported dissatisfaction reason, while the variant misses it and incorrectly attributes the reformulation to ``Not What I Wanted''. This case suggests that our reward shaping mechanism not only improves final satisfaction score prediction, as shown earlier, but also successfully enhances the quality of intermediate reasoning.
\vspace{-8pt}
\subsection{Quantitative Analysis}
\vspace{-14pt}
\begin{table}[H]
  \centering
  \caption{Counts and percentages of SearchLog26 test instances with 0, 1, or 2 reason categories shared by the predicted and user-reported sets.}
  \vspace{-10pt}
  \label{tab:reason-intersection}
  \small
  \begin{tabular*}{\linewidth}{@{}l@{\extracolsep{\fill}}rrrr@{}}
    \toprule
    & \multicolumn{3}{c}{
      \shortstack{Number of correctly predicted\\
      user-reported reason categories}
    } & \\
    \cmidrule(lr){2-4}
    Model & 0 & 1 & 2 & Total \\
    \midrule
    STaRS
      & 656 (49.77\%) & 631 (47.88\%) & 31 (2.35\%) & 1,318 \\
    w/o Reward Shaping
      & 790 (59.94\%) & 506 (38.39\%) & 22 (1.67\%) & 1,318 \\
    \bottomrule
  \end{tabular*}
  \vspace{-20pt}
\end{table}
\begin{table}[H]
  \centering
  \caption{Agreement between predicted and user-reported reason sets on the test set of the SearchLog26 dataset.}
  \vspace{-10pt}
  \label{tab:reason-set-similarity}
  \small
  \begin{tabular*}{\linewidth}{@{}l@{\extracolsep{\fill}}cc@{}}
    \toprule
    Model & Mean Jaccard (\%) $\uparrow$ & Mean F1 (\%) $\uparrow$ \\
    \midrule
    STaRS & \textbf{19.07} & \textbf{26.80} \\
    w/o Reward Shaping & 14.51 & 20.61 \\
    \bottomrule
  \end{tabular*}
  \vspace{-12pt}
\end{table}
We further quantitatively evaluate both models’ intermediate interpretations  following the same evaluation principle as in the case study across all 1,318 test instances in SearchLog26. Across test instances, the number of user-reported reasons ranges from 1 to 4 (mean: 1.50), with an illustrative reason set of [``Relevant'', ``Comprehensive'', ``Trustworthy''].
Specifically, for each instance \(i\) with \(N\) behavior signals, we first use GPT-5~\cite{singh2025openai}, which serves as a strong automatic category-mapping tool in this analysis, to map each natural-language interpretation \(\hat e_j\), together with its corresponding polarity \(\hat p_j\) (\(j=1,\ldots,N\)), to the best-fitting reason category shown in Figure~\ref{fig:app}, and then compare the resulting set of predicted reason categories with the set of user-reported reason categories for the same instance. 

For metrics, $G_{\mathrm{pred}}$ and
$G_{\mathrm{user}}$ denote the sets of predicted and
user-reported reason categories for one instance, respectively. The number of correctly predicted
user-reported reason categories is
$
\left|G_{\mathrm{pred}} \cap
G_{\mathrm{user}}\right|
$ for each instance.
$\text{Jaccard} =|G_{\mathrm{pred}}\cap G_{\mathrm{user}}|/|G_{\mathrm{pred}}\cup G_{\mathrm{user}}|$;
$F_1=2|G_{\mathrm{pred}}\cap G_{\mathrm{user}}|/(|G_{\mathrm{pred}}|+|G_{\mathrm{user}}|)$. Both metrics are computed for each instance. Table~\ref{tab:reason-set-similarity} reports the mean Jaccard and mean F1 scores across instances.

As shown in Table~\ref{tab:reason-intersection},
STaRS recovers at least one user-reported reason in
662 instances (50.23\%), compared with 528 (40.06\%) for
the variant without reward shaping, a gain of 10.17 percentage points.
Table~\ref{tab:reason-set-similarity} also shows higher mean Jaccard
(19.07\% vs.\ 14.51\%) and mean F1 (26.80\% vs.\ 20.61\%) over test instances,
indicating closer agreement with the full user-reported reason sets.
In summary, together with the satisfaction-prediction gains observed in the ablation study,
the qualitative and quantitative findings suggest that improved intermediate
interpretations may help explain the benefits of reward shaping for improved satisfaction
prediction performance. They also suggest potential training value for the behavior-signal-level reference supervision
generated by the high-capability LLM and filtered through our procedure, because, although we lack human-annotated behavior-level satisfaction polarities to directly assess the accuracy of the reference polarities through agreement with human judgments, experiments in this section provide indirect evidence of their training value for improving the accuracy of behavior-level polarity predictions: removing reference supervision reduces the agreement between the trained model’s interpretations and users’ reported reasons, which serves as a proxy for interpretation accuracy, and because the model derives its predicted polarities from these interpretations, this finding supports the value of the reference supervision in improving the accuracy of behavior-level polarity predictions.

\vspace{-8pt}
\section{Implementation Details of Baselines}
\label{sec:baseline}
\begin{table*}[h]
\centering
\small
\caption{Input features for baseline satisfaction prediction methods incorporating user behaviors. ``Trad. results'' is an abbreviation for ``traditionally ranked results''.}
\label{tab:features}
\vspace{-10pt}
\resizebox{\textwidth}{!}{
\begin{tabular}{lll}
\toprule
Group & Features & Description \\
\midrule

\multicolumn{3}{l}{\textbf{AI overview content semantic-level feature}} \\

AI overview response quality &
\texttt{AI\_Overview\_response\_quality\_score} &
A single normalized quality score for the entire AI overview response. \\

\midrule

\multicolumn{3}{l}{\textbf{AI overview examination behavior-level features}} \\

AI overview examination count&
\texttt{AI overview\_examination\_count} &
AI overview examination count. \\

AI overview examination time &
\texttt{AI overview\_examination\_dwell\_\{sum,min,max,mean\}} &
Examination time on the SERP for the AI overview. \\

\midrule

\multicolumn{3}{l}{\textbf{Trad. results examination and click behavior-level features}} \\

Trad. results examination count&
\texttt{trad\_examination\_count}&
Examinations of traditionally ranked results. \\

Trad. results click count&
\texttt{trad\_click\_count} &
Clicks on traditionally ranked results. \\

Trad. results examination time &
\texttt{trad\_serp\_dwell\_\{sum,min,max,mean\}} &
Examination time on the SERP for traditionally ranked results. \\

Trad. results content time &
\texttt{trad\_content\_dwell\_\{sum,min,max,mean\}} &
Dwell time on content pages for traditionally ranked results. \\

Exposed rank &
\texttt{exposed\_rank\_\{min,max,mean\}} &
Rank statistics of exposed results. \\

Clicked rank &
\texttt{clicked\_rank\_\{min,max,mean\}} &
Rank statistics of clicked results. \\

\midrule

\multicolumn{3}{l}{\textbf{Query reformulation behavior-level features}} \\

Query reformulation flag &
\texttt{has\_next\_query} &
Whether the user reformulates the query subsequently.\\

Query reformulation similarity&
\texttt{reformulation\_jaccard} &
Similarity between the next query and the current query.\\
\bottomrule
\end{tabular}
}
\end{table*}
For baselines in the\textbf{ first and second categories}, note that since ground-truth answers are unavailable in our datasets, we evaluate these baselines using only metrics that do not require ground-truth answers, so we exclude OpenEval~\cite{ispas2025towards} ’s correctness metric; we also cannot reproduce RAGCHECKER~\cite{3737916.3738608}, which requires a complete and correct ground-truth answer for each question. For baselines that produce scores on different scales,
we map their outputs to a common five-level satisfaction
scale.
For example, OpenEval~\cite{ispas2025towards} faithfulness scores in $0,1$
are discretized into five levels.

For traditional machine learning methods in the \textbf{third category} of baselines, to ensure a fair comparison with STaRS, as shown in Table~\ref{tab:features}, we extract a comprehensive set of input features that capture the same behavioral information used by STaRS.  In addition, for the semantic-level feature of the AI overview content, we use DeepSeek-V3.2 to assign a quality score to the entire AI overview response following Shi et al.~\cite{10.1145/3805712.3809934}. 
For sequence-based methods, including LSTM and Transformer, we adopt the same implementation as Shi et al.~\cite{10.1145/3805712.3809934}.
\vspace{-8pt}
\section{Impact of Reward Weights }
\begin{figure}[h]
\vspace{-10pt}
    \centering
    \includegraphics[width=1\linewidth]{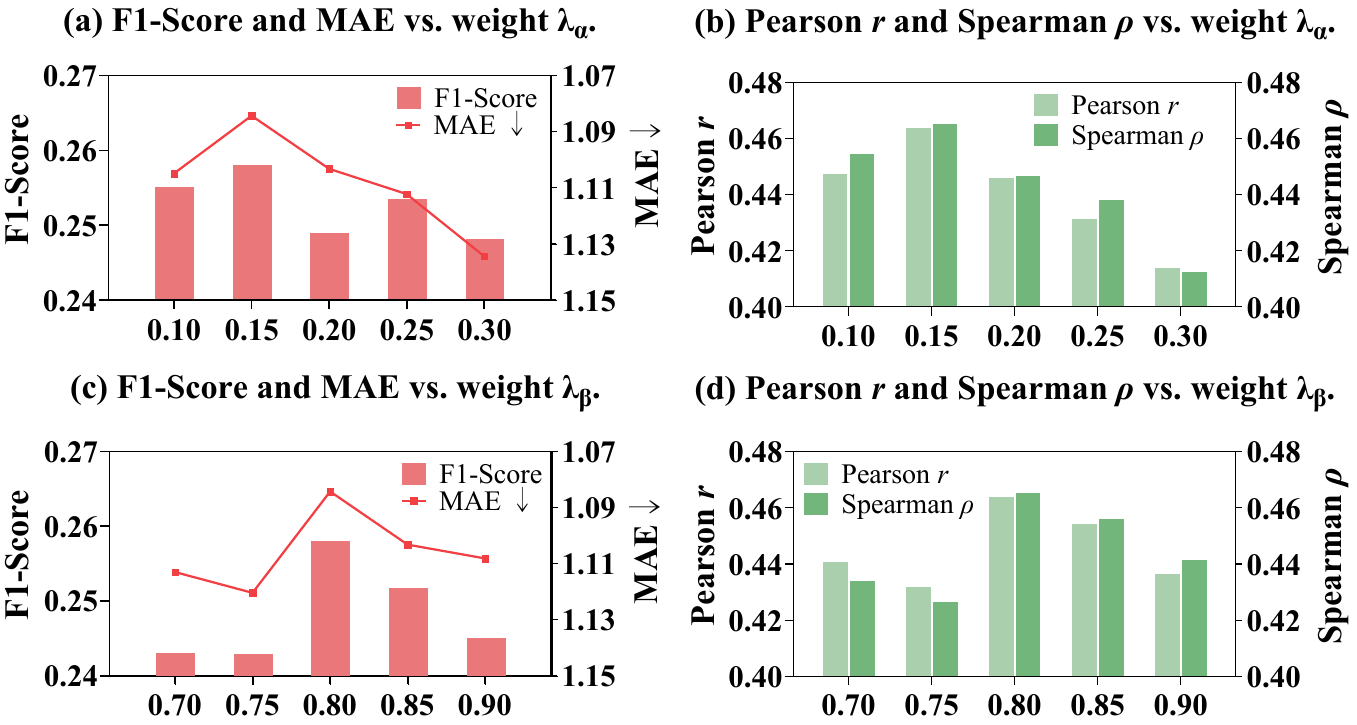}
    \vspace{-20pt}
    \caption{\textbf{Hyperparameter sensitivity analysis of reward weights for STaRS.} Each weight is varied while the other is held fixed. 
    }
    \vspace{-12pt}
    \label{fig:zhuzhuangtu}
\end{figure}
To understand how reward weights affect model performance, we perform sensitivity tests. We also use Qwen3-8B as the backbone of STaRS in this experiment. We use the same SearchLog26 train/test split as in the main experiment, with RL settings given in Appendix~\ref{sec:details_RL}.
We vary one reward weight at a time while keeping the other fixed. Specifically, in Figure~\ref{fig:zhuzhuangtu}, we fix $\lambda_{\beta}=0.8$ in panels (a, b) and $\lambda_{\alpha}=0.15$ in panels (c, d), varying the other weight over nearby values. As shown in Figure~\ref{fig:zhuzhuangtu}, panels (a, b) show the effect of weight \(\lambda_{\alpha}\), while panels (c, d) show the effect of weight \(\lambda_{\beta}\); panels (a, c) report F1-score and MAE, while panels (b, d) report Pearson \(r\) and Spearman \(\rho\).
The results suggest that STaRS achieves the best overall satisfaction prediction performance with $\lambda_{\alpha}=0.15$ and $\lambda_{\beta}=0.8$.
When $\lambda_{\beta}$ is too small or $\lambda_{\alpha}$ is too large, optimization may emphasize final predictions while providing insufficient guidance for interpreting individual behavior signals.
Conversely, an excessively large $\lambda_{\beta}/\lambda_{\alpha}$ ratio may cause the model to focus too heavily on individual behavior signals rather than effectively integrating them for overall satisfaction prediction, potentially degrading satisfaction prediction performance.
\vspace{-8pt}
\section{Full Prompt for the Structured Reasoning Strategy of STaRS}
\begin{table}[h]
\centering
\caption{Prompt used in our structured reasoning strategy for predicting user satisfaction with AI overviews.}
\vspace{-10pt}
\label{tab:structured_reasoning_prompt}
\footnotesize
\begin{tabular}{p{\dimexpr\linewidth-2\tabcolsep\relax}}
\hline

\#\#\# Task \newline
You are an expert in information retrieval evaluation. Given $( q,a,\{(b_i,c_i)\}_{i=1}^{N})$, predict the user's satisfaction score for the current AI overview on a five-point scale. \\
\hline

\#\#\# Instructions \newline
First, reason within \textless{}think\textgreater{} tags. Then, within \textless{}answer\textgreater{} tags, interpret each signal's satisfaction implications and assign a polarity based on its interpretation: + (supports satisfaction), - (supports dissatisfaction), or 0 (neutral). Finally, integrate these interpretations and polarities to predict overall satisfaction. Polarities indicate only the direction of evidence. Do not determine the score by summing polarities. \\
\hline

\#\#\# Output requirements \newline
Within \textless{}answer\textgreater{} tags, return a JSON array containing one entry for each behavior signal, preserving its original ID, followed by an entry containing the overall satisfaction reason and satisfaction score. Append \#\#\# and the same satisfaction score after the array. \\
\hline

\#\#\# Output format \newline
\textless{}think\textgreater{}\textless{}/think\textgreater{}\newline
\textless{}answer\textgreater{}[\newline
\hspace*{1em}\{\newline
\hspace*{2em}"behavior\_signal\_id": "B1",\newline
\hspace*{2em}"interpretation": "\{interpretation\_1\}",\newline
\hspace*{2em}"polarity": "\{polarity\_1\}"\newline
\hspace*{1em}\},\newline
\hspace*{1em}...,\newline
\hspace*{1em}\{\newline
\hspace*{2em}"behavior\_signal\_id": "BN",\newline
\hspace*{2em}"interpretation": "\{interpretation\_N\}",\newline
\hspace*{2em}"polarity": "\{polarity\_N\}"\newline
\hspace*{1em}\},\newline
\hspace*{1em}\{\newline
\hspace*{2em}"reason": "\{reason\}",\newline
\hspace*{2em}"satisfaction\_score": \{satisfaction\_score\}\newline
\hspace*{1em}\}\newline
]\#\#\#\{satisfaction\_score\}\textless{}/answer\textgreater{} \\
\hline

\end{tabular}
\vspace{-12pt}
\end{table}
The prompt used in our structured reasoning strategy is shown in Table~\ref{tab:structured_reasoning_prompt}. The model receives $( q, a,\{(b_i,c_i)\}_{i=1}^{N})$ as input. The behavior signals are ordered chronologically by the start time of the earliest behavior event represented by each signal and assigned unique behavior signal IDs. The model is instructed to generate a structured JSON output enclosed in \texttt{<answer>} tags.
\vspace{-8pt}
\section{Additional Details of RL Training and Reward Shaping of STaRS}
\label{sec:details_RL}
We consistently adopt the VERL~\cite{10.1145/3689031.3696075} framework for RL of open-source LLMs and perform training on 4 NVIDIA A800 80GB GPUs. We train the policy model for 5 epochs with a learning rate of \(1 \times 10^{-6}\). We incorporate KL regularization into the policy loss. We employ vLLM for acceleration. 

When reference supervision is required, we use DeepSeek-V3.2 to provide offline reference supervision, with $K=5$ candidate traces per training instance. When counterfactual signal selection is enabled, we use DeepSeek-V3.2 to re-estimate satisfaction, with each re-estimation performed $Z=3$ times.

\end{document}